\documentclass[aps,pre,twocolumn,letterpaper,showpacs,superscriptaddress,preprintnumbers,amsmath,amssymb,floatfix,nofootinbib,showkeys]{revtex4-1}
\pdfoutput=1
\usepackage{graphicx}
\usepackage{epsfig,color}
\usepackage{bbm,amsmath,amssymb,psfrag}
\usepackage{dcolumn}
\usepackage[utf8x]{inputenc}
\usepackage{MnSymbol}
\usepackage[T1]{fontenc}
\usepackage{url}
\usepackage{bm}  
\usepackage[color]{changebar} 
\usepackage{cases}

\def\W{\mathbf{W}}
\def\H{\mathbf{H}}
\def\U{\mathbf{U}}
\def\V{\mathbf{V}}

\def\Q{\mathbf{Q}}
\def\R{\mathbf{R}}
\def\x{\mathbf{x}}
\def\y{\mathbf{y}}
\def\u{\mathbf{u}}
\def\v{\mathbf{v}}
\def\h{\mathbf{h}}

\def\f{\mathbf{f}}
\def\g{\mathbf{g}}

\def\P{\mathcal{P}}

\def\E{\mathcal{E}}
\def\O{\mathcal{O}}
\def \uba{\u_{\backslash a}}

\def \ubah{\hat{\u}_{\backslash a}}
\def \ubap{\u^\prime_{\backslash a}}
\def \T{\mathrm{T}}
\def \chiba{\boldsymbol{\chi}_{\backslash a}}
\def \chibai{\boldsymbol{\chi}_{\backslash a i}}
\def \seff {\sigma_\mathrm{eff}^2}
\def \sxi{\sigma_\xi^2}
\def\CHI{\boldsymbol{\chi}}

\begin{document}

\title{The cavity method for analysis of  large-scale penalized regression}

\author{Mohammad Ramezanali}
\email{mrr@physics.rutgers.edu}
\affiliation{Department of Physics and Astronomy, Rutgers University, 136 Frelinghuysen Rd, Piscataway, NJ 08854 USA}
\author{Partha P. Mitra} 
\email{mitra@cshl.edu}
\affiliation{Cold Spring Harbor Laboratory, 1 Bungtown Road, Cold Spring Harbor, NY 11734 USA}
\author{Anirvan M. Sengupta}
\email{anirvans@physics.rutgers.edu}
\affiliation{Department of Physics and Astronomy, Rutgers University, 136 Frelinghuysen Rd, Piscataway, NJ 08854 USA}
\affiliation{Center for Quantitative Biology, Rutgers University, 110 Frelinghuysen Rd, Piscataway, NJ 08854 USA}


\begin{abstract}

Penalized regression methods aim to retrieve reliable predictors among a large set of putative ones from a limited amount of  measurements. In particular, penalized regression with singular penalty functions is important for sparse reconstruction algorithms. For large-scale problems, these algorithms exhibit sharp phase transition boundaries where sparse retrieval breaks down.  Large optimization problems associated with sparse reconstruction have been analyzed in the literature by setting up corresponding statistical mechanical models at a finite temperature. Using replica method for mean field approximation, and subsequently taking a zero temperature limit, this approach reproduces the algorithmic phase transition boundaries.  Unfortunately, the replica trick and the non-trivial zero temperature limit obscure the underlying reasons for the failure of a sparse reconstruction algorithm, and of penalized regression methods, in general. In this paper, we employ the ``cavity method'' to give an alternative derivation of the mean field equations, working directly in the zero-temperature limit. This derivation provides insight into the origin of the different terms in the self-consistency conditions. The cavity method naturally involves a quantity, the average local susceptibility, whose behavior distinguishes different phases in this system. This susceptibility can be generalized for analysis of a broader class of sparse reconstruction algorithms.

\end{abstract}



\maketitle


\section{Introduction}

In traditional statistics, we are given a sample of random observations much larger than the number of unknown parameters being estimated. However, during the last couple of decades, data collection has become much more automatic and much more extensive. As a result, the limit of large number of unknown parameters appears in a variety of fields, ranging from communication technology to business informatics to systems biology, posing challenges to the classical statistical paradigm. 
Many methods for the selection of important variables, proposed within the statistics literature,  are combinatorial in nature~\cite{weisberg2005applied}.  The explosion of number of possibilities, when the number of unknown variables are comparable to the sample size, places a huge computational burden on the more principled  methods, severely limiting their applicability to large data sets. Practical variable selection procedures drastically limit the search space, but they are often greedy in nature and potentially unreliable.

Variable selection via penalized regression~\cite{Tibshirani96, hoerl1962application} has gained wide appeal, owing to these concerns regarding combinatorial methods.  One common setup is a linear model, $\y=\H\x_0+\bm\zeta$, where it is assumed that $\H$ is a known ${M \times N}$ measurement matrix,  $\x_0$ is an unknown $N$-dimensional signal vector, possibly sparse, and the vector $\zeta$ models the measurement noise. The task is to reconstruct $\x_0$ from $\y$ and $\H$, utilizing, in principle, the knowledge about the statistics of the signal and of noise.

The general form of penalized methods for standard linear model, in practice, does not require detailed information about statistics of $\x_0$ and $\zeta$. The task reduces to minimizing the following objective function
\begin{equation}
\hat{\x}(\lambda\sigma^2)= \underset{\x}{\mathrm{arg\,min}} \{\frac{1}{2\sigma^2} \left( \y-\H\x\right )^2 +  \lambda V(\x)\}.
\label{eq:LS}
\end{equation}
where $\lambda\sigma^2$ is a non-negative parameter deciding relative weight between quadratic loss function and the penalty function $V$ in Eq.~\eqref{eq:LS}.  The penalty function $V$ is chosen to suppress effects of noise on estimation as well as to regularize potentially `soft' directions in the loss function. A popular class of penalties are of the form $\sum_a |x_a|^m.$  The choice $m=2$ or $\ell_2$ penalty is known as Ridge regression~\cite{Tikhonov43}, a convex penalty function  which has been shown to give more accurate predications, when sparsity of the solution is not important. Choosing $m\to0$ yields the so-called $\ell_0$ penalty which controls sparsity directly but leads to a non-convex optimization problem. Finally, $m=1$ or $\ell_1$ penalty, has the benefits of being a convex function as well as of promoting sparsity~\cite{Tibshirani96, Chen98}.

When the vector $\x$ has $K$ non-zero components, with  $K \leq M$ and  $M \leq N$, there are many rigorous results on the performance of sparse reconstruction based on $\ell_1$-penalization~\cite{CandesCS,Donoho06}. As a result of the computational attractiveness of convex optimization and of the performance guarantees  which reach optimality under certain conditions, $\ell_1$-penalization has become a popular method, particularly in the context of so-called Compressed Sensing~\cite{Donoho06}.  In the asymptotic limit $M, N, K \to \infty$, there is a phase transition in noiseless sparse recovery~\cite{DonohoPT, donoho2006thresholds}, with constrained optimization of the $\ell_1$ penalty leading to perfect reconstruction in the `good' phase. 

As one can imagine,  the statistical physics of disordered systems offers powerful tools to understand this phase transition in the asymptotic limit. In a series of works using the {\it replica method}, several investigators have studied the performance of $\ell_1$-penalized regression and the algorithmic phase transition~\cite{Rangan12, Guo05,Tanaka02, Kabashima09, Ganguli10}. In comparison to rigorous results providing guarantees~\cite{CandesCS}, the statistical mechanics approach enables a detailed analysis of the behavior of the optimization algorithms near the region of failure. On the other hand, the replica trick  is applied to a finite temperature system. The non-trivial zero temperature limit hides the essential role of local susceptibility in deciding the robustness of performance of the algorithms.  Also, the derivation of the self-consistency conditions obscures the subtlety of certain aspects of this mean field theory.

An alternative to the replica trick, with the number of replicas `mysteriously' tending to zero, is to use the {\it cavity method}~\cite{MPCavity, MPBook}. It is a direct approach to mean field theory, originally designed for understanding the nature of ground states of certain spin glass models. The method has since been applied to a wider class of problems including algorithmic phase transitions, some examples being the satisfiability problem~\cite{MuletCol,MezardSAT} and Hopfield neural networks~\cite{Sompolinsky00}. The cavity method leads to the same results as obtained by replica trick~\cite{MPBook} for spin glass mean field theory  and is closely related to the message-passing algorithm in graphical models~\cite{Message-Passing-Graph}. We find that for the problem at hand, the cavity method provides better insight in comparison to the replica formalism and  has the potential to lead to substantially better sparse reconstruction algorithms.

The layout of this paper is as follows. In the next two sections, we briefly review a finite noise/finite temperature formulation of the problem and the replica approach to the mean field theory, facilitating comparison with our analysis via the cavity method. Readers familiar with these results, or those solely interested in the cavity method, could skim through these sections just familiarizing themselves with the notation. In Sec.~\ref{sec:result}, we introduce a susceptibility matrix associated with this problem and then provide an outline of the derivation of  the self-consistent mean field equations via a two-step cavity method working directly at zero temperature. This approach is not only different from the replica method but also from the analysis based on iterations in a message-passing algorithm~\cite{Montanari-book, DonohoAMP}. We end by illustrating how our cavity mean field picture relate to success and to failure of sparse reconstruction in medium sized $\ell_1$-penalized  problems in Sec.~\ref{sec:numeric}. The appendix has additional technical details of the derivation, along with the cavity approach worked out for finite temperature.


\section{Statistical Mechanics Formulation}
\label{sec:reconstruction}

Here, we set up the general framework for investigating the regularized least-squares reconstruction algorithms. We assume that the data $\y=\H\x_0+\bm\zeta$ are generated by a probability distribution $p(\y|\x_0,H)$, given an (unknown) sparse signal $\x_0$ and a (known) matrix $\H$, and an (unknown) Gaussian noise vector $\bm\zeta$ whose components are i.i.d. samples from $\mathcal{N}(0,\sigma_\zeta^2)$. The vector  $\x_0$ is considered to be a random sample from a distribution $P_0(\x_0)=\prod_a p_0(x_{a0})$. 


Although, in general,  the probability distribution of $\H$, $\mathcal{P}(\H)$, could be a non-Gaussian distribution, at this point we consider it to be a multivariate Gaussian distribution.  The element-wise mean and the covariance matrix entries are given by
\begin{equation}
\big[H_{i a}\big]^{\mathrm{av}} = 0
\label{eq:gauss-av}
\end{equation} 
\begin{equation}
\big[ H_{i a} H_{j b}]^{\mathrm{av}} = \frac{1}{M}\delta_{i j}\delta_{a b}.
\label{eq:gauss-covar}
\end{equation} 

We study the performance of an estimator of $\x_0$, namely the location $\hat \x$ of the minimum of a cost function 
\begin{equation}
\E_0(\x)=\frac{(\y-\H\x )^2}{2 \sigma^2}+ V(\x). 
\label{eq:cost}
\end{equation}
We can reformulate the cost minimization as a statistical mechanics problem where the cost function will play the role of energy. We assume the penalty/potential term $V(\x)$ is such that there is a unique minimum of $\E_0$.
Note that $\hat \x={\mathrm{arg\,min}}_x\!\ \E_0(\x)$ depends on $\y,\H$, meaning that it can be written as a function  $\hat \x=\bm g(\x_0,\H, \bm\zeta)$, using the fact that $\y=\H\x_0+\bm\zeta$. We have set up an ensemble of problem instances by specifying the probability distribution of the variables $\x_0,\H, \bm\zeta$, so that we could study the performance of the estimator over this ensemble. In particular, it will be useful to extract moments of the distribution of the parameter estimation error $\hat \x-\x_0$.

In order to make a connection between the optimizations problem and statistical mechanics,  one could choose a probability distribution of $\x$ parametrized by $\beta$, playing the role of inverse temperature,
\begin{align}
p_\beta(\x|\y,\H)=&\frac{1}{Z}\mathrm{exp} \big(-\beta\E_0(\x)\big)\nonumber\\
=&\frac{1}{Z(\beta,\y,\H)}\mathrm{exp} \Bigg\{-\beta\bigg(\frac{(\y-\H \x)^2}{ 2 \sigma^2}+V(\x)\bigg)\Bigg\}
\label{eq:pbeta}
\end{align}
with the normalization factor $Z=Z(\beta,\y,\H)$, known as the partition function,  given by
\begin{align}
Z (\beta,\y,\H)=&\int d^N\x\frac{1}{Z}\mathrm{exp} \big(-\beta\E_0(\x)\big)\nonumber\\
=&\int d^N\x  \mathrm{exp} \Bigg\{-\beta\bigg(\frac{(\y-\H \x)^2}{ 2 \sigma^2}+V(\x)\bigg)\Bigg\}. 
\label{eq:partition}
\end{align}
If we send $\beta$ to $\infty$, equivalent to sending the temperature to zero, the probability gets concentrated at the minimum of the cost/energy function. Keep in mind that we define $\beta$ to be dimensionless. 

We will consider averages of functions of the form $\O(\x,\x_0)$ containing both the original sparse signal and the variable related to the estimate.   The average of the function $\O(\x,\x_0)$  over the distribution $p_\beta(\x|\y,\H)$ is given by
\begin{equation}
\langle \O(\x,\x_0)\rangle = \frac{\int d^N\x \O(\x,\x_0)\ \mathrm{exp}\{-\beta\frac{(\y-\H\x )^2}{2 \sigma^2}-\beta V(\x)\}}{\int d^N\x  \mathrm{exp}\{-\beta\frac{(\y-\H\x )^2}{2 \sigma^2}-\beta V(\x)\}}.
\label{eq:func-expect-expand}
\end{equation}
This `thermal' average, represented by $<\cdots>$, depends on the random variables $\x_{0}$, $\H$ and $\bm \zeta$.  Note that in the limit  $\beta\rightarrow\infty$, this average should become $f(\hat \x,\x_0)$, for continuous $f$. Averaging the result of this calculation over the random instances of $\x_{0}$, $\H$ and $\bm \zeta$ is a technical challenge related to quenched averages in disordered systems \cite{MPBook}.

The function $\O(\x,\x_0)=\frac{1}{N}(\x-\x_0)^2$ plays an important role in our analysis. Its average corresponds to the mean squared estimation error:
\begin{equation}
{\rm MSE}\equiv \frac{1}{N} \sum_{a=1}^N \big[\langle x_a-x_{a0}\rangle^2\big]^{\mathrm{av}}_{\x_0,\H,\bm\zeta}= \frac{1}{N}\big[\langle(\x-\x_0)^2\rangle\big]^{\mathrm{av}}_{\x_0,\H,\bm\zeta}.
\label{eq:mse}
\end{equation} 
We will use $\big[\cdots\big]^{\mathrm{av}}_{\rm vars}$ to denote quenched averages, with the relevant quenched variables indicated in the subscript, when necessary.
We use the notation $\u=\x-\x_0$ to indicate the estimation error vector. 
The size of the vector $\u$ provides a measure of the inaccuracy of the reconstruction.

In the context of penalized regression, the penalty function is often chosen to be a sum of potentials involving single variables, namely, $V(\x)=\sum_aU(x_a)$. We will focus on $V(\x)$ of this nature. An important special case for example is in compressed sensing with sparsity promoting regularizing potential $U(x)=\lambda |x|$. In this paper, we would mostly restrict ourselves to the noiseless case, $\bm\zeta=0$, although the same methods could be used to analyze the noisy case as well.

For $\bm \zeta=0$, we will be interested in the result of the constrained optimization problem of minimizing $V(\x)$ subject to the constraint $\y=\H\x$.  In the $M,N\rightarrow \infty$ limit, this problem may exhibit a phase transition from a perfect reconstruction phase to  an error-prone phase, with the MSE, mentioned above, as the order parameter. This constrained optimization could be studied in more than one equivalent ways. After taking  $\beta \rightarrow \infty$ limit, we will take the route $\sigma\rightarrow 0$ to enforce the equality $\y=\H\x$. 


\section{Replica Approach}
\label{sec:replica}
In this section, we review the replica approach to the problem~\cite{Kabashima09,Ganguli10}, presenting the mean field equations in terms of a distribution of asymptotically independent single-variable problems with a set of self-consistency conditions.
In order to calculate quantities like the MSE, we need to compute quenched averages of the form $\big[\langle \O(\x,\x_0)\rangle\big]^{\mathrm{av}}$, which is complicated by the presence of the denominator in  Eq.~\eqref{eq:func-expect-expand}.  Formally, the denominator is handled by introducing $n$ non-interacting replicas of the system and taking $n\rightarrow 0$, as shown below.  In the noiseless case, $\E_0(\x)$ depends on $\x$ as well as on $\x_0,H$. To emphasize those additional dependences, we write $\E_0(\x)$ as $\E_0(\x_\mu,\x_0,\H)$ in the next few equations.
\begin{align}
\langle& \O(\x,\x_0)\rangle_{\x} = 
 \frac{\int d^N\x \O(\x,\x_0)\exp \big(-\beta \E_0(\x,\x_0,\H)\big)}
{\int d^N\x\exp \big(-\beta \E_0(\x,\x_0,\H)\big)}\nonumber\\
 =& \lim_{n  \rightarrow 0}\bigg(\int d^N\x\exp \big(-\beta \E_0(\x,\x_0,\H)\big)\bigg)^{n-1}\nonumber\\
& \int d^N\x \O(\x,\x_0)\exp \big(-\beta \E_0(\x,\x_0,\H)\big)\nonumber\\
=&\lim_{n  \rightarrow 0}\int \O(\x_1,\x_0)\prod\limits_{\mu=1}^n \big\{d^N\x_\mu\exp \big(-\beta \E_0(\x_\mu,\x_0,\H)\big)\big\}.
 \label{eq:replica-for-av}
\end{align}
Averaging over the quenched variables $\x_0$ and $H$, we get 
\begin{align}
 &\big[ \langle \O(\x,\x_0) \rangle_{\x} \big]^{\mathrm{av}}_{\x_0,\H} \nonumber\\
 &=  \lim_{n  \rightarrow 0}\bigg[ \int\prod\limits_{\mu=1}^n \{d^N\x_\mu\} \O(\x_1,\x_0)\exp \big(-\beta\sum\limits_{\mu} \E_0(\x_\mu,\x_0,\H)\big)\bigg]^{\mathrm{av}}_{\x_0,\H}.
 \label{eq:quenched-av}
\end{align}
Using $\y=\H\x_0$ in the noiseless case, the energy function for the $n$-th replica would be
\begin{align}
&\E_0(\x_\mu,\x_0,\H)=\frac{(\y-\H\x_\mu )^2}{2 \sigma^2}+ V(\x_\mu)\nonumber\\
&=\frac{(\H\x_0-\H\x_\mu )^2}{2 \sigma^2}+ V(\x_\mu)
=\frac{\H\u_\mu ^2}{2 \sigma^2}+ V(\u_\mu+\x_0), 
\label{eq:cost-replica}
\end{align}
rewritten in terms of the error variables $\u_{\mu} = \x_\mu-\x_0, \mu = 1,..,n$.
Thus, we are interested in average quantities in the replicated ensemble whose partition functions is given by
\begin{align}
&\big[{Z^n}\big]^{\mathrm{av}}_{\x_0,\H} \nonumber\\
 =&\bigg[  {\int}  \prod_{\mu=1}^n \, d\u_\mu \exp\big[- \beta\big\{\sum_{\mu=1}^n \frac{\left ( \H \u_{\mu} \right )^2}{2\sigma^2}+V(\u_{\mu}+\x_0)\big\}\big]\bigg]^{\mathrm{av}}_{\x_0,\H}.
\label{eq:partition-func-expand}
\end{align}
In order to average over $\mathcal P(\H)$, the only quantity that needs to be computed is 
\begin{align}
&\bigg[  \exp\big(- \sum_{\mu=1}^n \frac{\beta}{2\sigma^2}\left ( \H \u_{\mu} \right )^2\big)\bigg]^{\mathrm{av}}_{\H}\nonumber\\
=&\frac{1}{Z_0}  {\int}  d\H \, \mathrm{exp}\big[- \frac{M}{2}\mathrm{Tr}\left ( \H^{\bm\top} \H   \right )
 -\frac{\beta}{2\sigma^2}\sum_{\mu=1}^n \u_{\mu}^{\bm\top} \H^{\bm\top} \H \u_{\mu}\big]\nonumber\\
=& \bigg[\frac{1}{\det\big( \mathbf{I}_{ N}+\frac{\beta}{\alpha\sigma^2}\sum_{\mu=1}^n \frac{1}{N}\u_{\mu}\u_{\mu}^{\bm\top} \big)}\bigg]^{-M/2}
\label{eq:int-over-H}
\end{align}
where $Z_0$ is the normalization term for the Gaussian distribution of $\H$, and $\alpha=M/N$ is the sampling ratio. 
We can look at the $\phi=\bm{\mathcal{U} } \bm{\mathcal{U} }^{\bm\top} $ where $\bm{\mathcal{U} }$ is an $N$ by $n$ matrix, $[\u_1, \u_2,\cdots, \u_\mu]$. We write the singular value decomposition (SVD) for $\bm{\mathcal{U} }$ as 
\begin{equation}
\bm{\mathcal{U} }=\U\Lambda\V^\T
\end{equation}
where $\Lambda$ denotes the $n$ by $n$ matrix with $n$ non-zero singular values equals $\lambda_\mu$. Therefore, the
eigenvalues of $\phi$ are simply the square of these singular values. Changing the variable from $\u$ in \eqref{eq:int-over-H} to $\U$, $\V$, and $\Lambda$:
\begin{equation}
\det\big( \mathbf{I}_{ N}+\frac{\beta}{\alpha\sigma^2}\frac{1}{N}\bm{\mathcal{U} }\bm{\mathcal{U} }^{\bm\top} \big)
=\underset{\mu=1}{\prod^n} \big( 1+\frac{\beta}{\alpha\sigma^2}\frac{1}{N}\lambda_\mu^2\big).
\label{eq:int-over-H-svd}
\end{equation}
The right hand side of Eq. \eqref{eq:int-over-H-svd} can be written as $\det\big( \mathbf{I}_{ n}+\frac{\beta}{\alpha\sigma^2}\Q\big)$ with the elements of the $n\times n$ matrix $\Q$ are defined by $\Q = \tfrac{1}{N}\bm{\mathcal{U} }^{\bm\top}\bm{\mathcal{U} }$ which have the same $\lambda_\mu^2$ eigenvalues. Therefore, we can rewrite Eq. \eqref{eq:partition-func-expand} as

\begin{align}
&\big[{Z^n}\big]^{\mathrm{av}}_{\x_0,\H} \nonumber\\
 =&\bigg[  {\int}  \prod_{\mu=1}^n \, d\u_\mu \exp
 \bigg[-\frac{M}{2}\mathrm{Tr}\log\big( \mathbf{I}_{ n}+\frac{\beta}{\alpha\sigma^2}\Q\big)+\underset{\mu=1}{\sum^n}V(\u_{\mu}+\x_0)\bigg].
\label{eq:partition-func-expand-Q}
\end{align}
Using the Fourier representation of the $\delta$ function 
\begin{equation}
\delta ( \u_\mu^\top  \u_\nu - NQ_{\mu \nu} )  = \frac{1}{2\pi}\int dR_{\mu \nu} \mathrm{exp}\big(-iR_{\mu \nu} ( \u_\mu^{\bm\top}  \u_\nu - NQ_{\mu \nu} ) \big)
\label{eq:delta-func}
\end{equation}
and inserting this delta function with an integral over $\Q_{\mu \nu}$ in Eq.~\eqref{eq:partition-func-expand}, we get
\begin{align}
&\big[{Z^n}\big]^{\mathrm{av}}_{\x_0,\H} =  {\int}  \prod_{\mu\le\nu} dQ_{\mu \nu} dR_{\mu \nu}\exp[-S(\Q,\R)]\\ 
 &S[\Q,\R]=
\frac{M}{2}\mathrm{Tr}\log\big( \mathbf{I}_{ n}+\frac{\beta}{\alpha\sigma^2}\Q\big) 
-iN\mathrm{Tr}(\R\Q)\nonumber\\
&-\log\bigg[\int  \prod_{\mu=1}^n d\u_{\mu}\exp\big[-i\sum_{\mu,\nu}R_{\mu \nu} \u_{\mu}^{\bm\top}  \u_\nu+\sum_\mu V(\u_{\mu}+\x_0)
\big]\bigg]^{\mathrm{av}}_{\x_0}
\end{align}
This integral over $\Q,\R$ can be evaluated using the saddle point method~\cite{Kabashima09,Ganguli10} when $M,N  \rightarrow \infty$, holding $\alpha = \frac{M}{N}$ fixed. The saddle point $\Q=\bar{\Q}, \R=-i\bar{\R}$ satisfies the conditions:
\begin{equation}
\bar{Q}_{\mu \nu} = \frac{1}{N}\langle\langle \u_\mu^{\bm\top} \u_\nu \rangle\rangle
\label{eq:Q}
\end{equation}
\begin{equation}
\bar{\R} = \frac{\beta}{2\sigma^2}\big[\mathbf{I}_{ n}+\frac{\beta}{\alpha\sigma^2}\Q \big]^{-1}
\label{eq:R}
\end{equation}
obtained by differentiating $S(\Q,\R)$ with respect to the elements of $\Q,\R$. The expectation $\langle\langle \u_\mu^{\bm\top} \u_\nu \rangle\rangle$ depends on $\bar{\R}$ via

\begin{align}
&\langle\langle \u_\mu^{\bm\top} \u_\nu \rangle\rangle=\beta\frac{\partial F(\bar{\R})}{\partial \bar{R}_{\mu\nu}}\\
&\text{with } \exp\{-\beta F(\bar{\R})\}\nonumber\\
&=\bigg[\int  \prod_{\mu=1}^n\{d^N\u_\mu\} \exp\big[-\sum_{\mu,\nu}\bar{R}_{\mu \nu} \u_{\mu}^{\bm\top}  \u_\nu-\beta\sum_\mu V(\u_{\mu}+\x_0)
\big]\bigg]^{\mathrm{av}}_{\x_0}.
\label{eq:F}
\end{align}

If $U(x)$ is a convex function, we expect a unique state and a replica symmetric solution for $\Q,\R$. This implies $\bar{Q}_{\mu \nu} = (Q-q) \delta_{\mu \nu} + q$ and  $\bar{R}_{\mu \nu} =  (R-r) \delta_{\mu \nu} + r$. With that ansatz,
\begin{align}
& \int  \prod_{\mu=1}^n\{d^N\u_\mu\} \exp\big[-\sum_{\mu,\nu}\bar{R}_{\mu \nu} \u_{\mu}^{\bm\top}  \u_\nu-\beta\sum_\mu V(\u_{\mu}+\x_0)
\big]\nonumber\\
=&\int  \prod_{\mu=1}^n\{d^N\u_\mu\} \exp\big[-(R-r)\sum_\mu\u_\mu^2\nonumber\\
&-r(\sum_\mu \u_\mu)^2-\beta\sum_\mu V(\u_{\mu}+\x_0)
\big]\nonumber\\
=&\int \frac{d^N\bm\xi}{(2\pi\sxi )^{N/2}} \exp(-\frac{\bm\xi^2}{2\sxi})\int  \prod_{\mu=1}^n\{d^N\u_\mu\} \nonumber\\
&\exp\big[-\frac{\beta}{2\seff}\sum_\mu\u_\mu^2
+\frac{\beta}{\seff}\bm\xi^{\bm\top}(\sum_\mu \u_\mu)-\beta\sum_\mu V(\u_{\mu}+\x_0)
\big]
\label{eq:replica-symmetric}
\end{align}
identifying $R-r\equiv\tfrac{\beta}{2\seff}$ and $r\equiv-\tfrac{\beta^2\sxi}{2\sigma_\mathrm{eff}^4}$. We have used 
\begin{align}
&\int \frac{d^N\bm\xi}{(2\pi\sxi )^{N/2}} \exp(-\frac{\bm\xi^2}{2\sxi})
\exp[\frac{\beta}{\seff}\bm\xi^{\bm\top}(\sum_\mu \u_\mu)]\nonumber\\
&=\exp[\frac{\beta^2\sxi}{2\sigma_\mathrm{eff}^4}(\sum_\mu \u_\mu)^2]
\end{align}
to decouple the item replica coupling in the $(\sum_\mu \u_\mu)^2$ term, at the cost of introducing another quenched variable $\bm \xi$.
Note that we require $R-r>0$ and $r<0$ for this approach to work. These inequalities follow from~\eqref{eq:R} and from $Q-q>0$ and $q>0$. The conditions on $Q$ and $q$ would be obvious once we look at interpretation of these quantities described below.

For $V(\x)=\sum_aU(x_a)$ we can simplify further. Remembering that we also need to do the quenched average over $\x_0$,
\begin{align}
& \Bigg[\int  \prod_{\mu=1}^n\{d^N\u_\mu\} \exp\big[-\sum_{\mu,\nu}\bar{R}_{\mu \nu} \u_{\mu}^{\bm\top}  \u_\nu-\beta\sum_\mu V(\u_{\mu}+\x_0)\big]\Bigg]^{\mathrm{av}}_{\x_0}
\nonumber\\
=&\Bigg[\int  \prod_{\mu=1}^n\{d^N\u_\mu\} 
\exp\bigg[-\beta\bigg\{\frac{1}{2\seff}\sum_\mu(\u_\mu^2-\bm\xi^{\bm\top} \u_\mu)\nonumber\\
&\qquad\qquad\qquad\qquad\qquad+\sum_\mu V(\u_{\mu}+\x_0)\bigg\}
\bigg]\Bigg]^{\mathrm{av}}_{\bm\xi,\x_0}\nonumber\\
=&\prod_a\Bigg[\int  \prod_{\mu=1}^n\{du_{\mu a}\} 
\exp\bigg[-\beta\bigg\{\frac{1}{2\seff}\sum_\mu(u_{\mu a}^2-\xi_a u_{\mu a})\nonumber\\
&\qquad\qquad\qquad\qquad\qquad+\sum_\mu U(u_{\mu a}+x_{0a})\bigg\}
\bigg]\Bigg]^{\mathrm{av}}_{\xi_a,x_{0a}}
\label{eq:replica-symmetric-factorized}
\end{align}
Thus, in the saddle point approximation, each of the $N$ components of $\u$ become effectively independent and the saddle point conditions reduce to a self-consistent problem for each component $a=1,\ldots,N$. Since this self-consistent problem is similar for each index,  we suppress the subscript $a$ in $u_{\mu a}$ and in $x_{0a}$. For each $a$, we have the integral of the form
\begin{equation}
\Bigg[\int  \prod_{\mu=1}^n du_\mu
\exp\bigg[-\beta\bigg\{\frac{1}{2\seff}\sum_\mu(u_{\mu }^2-\xi u_{\mu })\nonumber\\
+\sum_\mu U(u_{\mu }+x_{0})\bigg\}
\bigg]\Bigg]^{\mathrm{av}}_{\xi,x_0}.
\label{eq:replica-symmetric-single-variable}
\end{equation}

The replica problem corresponds to a single variable $u$ follows the effective distribution 
\begin{equation}
P_\mathrm{eff}(u \, | x_0,\xi) = \frac{1}{Z(x_0,\xi)} e^{-\beta\E_\mathrm{eff}(u;x_0,\xi)},
\label{eq:prob-eff}
\end{equation} 
with an effective mean-field Hamiltonian
\begin{equation}
\E_\mathrm{eff}(u;x_0,\xi)=\frac{1}{2\seff} \left (u^2 -2\xi u\right )+  U(u+x_0)
\label{eq:Eeff-replica}
\end{equation}
which depends on two quenched variables $x_0$ and $\xi$. The variable $x_0$ has the probability distribution $p_0(x_0)$, whereas $\xi$ is distributed according to a Gaussian distribution with mean zero and variance $\sxi$. The two parameters $\seff$ and $\sxi$  are given by the following set of self-consistency conditions

\begin{equation}
q = [ \langle u\rangle^2]^{\mathrm{av}}_{x_0,\xi}, \quad
\Delta Q\equiv Q-q =[ \langle (u-\langle u\rangle)^2\rangle]^{\mathrm{av}}_{x_0,\xi}
\label{eq:dQ}
\end{equation}
\begin{equation}
\seff = \sigma^2+\frac{\beta\Delta Q}{\alpha}, \quad
\sxi=\frac{q}{\alpha}
\label{eq:sigeff}
\end{equation}
where the thermal averages $\langle{\cdots}\rangle$ over $u$ are performed in the $P_\mathrm{eff}$ ensemble and the so-called quenched average $[ \cdots ]^{\mathrm{av}}_{x_0, \xi}$ is over variables $\xi$ and $x_0$. 

In order to study the regularized least-squares reconstruction, we need to take the limits $\beta\rightarrow\infty$, and then $\sigma\rightarrow0$. A nontrivial aspect of the zero temperature limit ($\beta\rightarrow\infty$) is the quantity $\beta\Delta Q$ in Eq.~\eqref{eq:sigeff} that behaves differently in different phases of reconstruction. Using Eq.~\eqref{eq:dQ}, this quantity is just $\beta$ times the thermal fluctuation in $u$. The fluctuation-dissipation relation~\cite{Kubo66} implies that this quantity may be interpreted as a local susceptibility. 

In our following considerations based on the zero temperature cavity method, we formally introduce a susceptibility and use its properties to give a more transparent derivation of the same equations via two step cavity method.
We will outline the main points of the derivations in the next section, and refer the reader to Appendices \ref{app:susc} and \ref{app:cav} for the details.


\section{Outline of the Cavity Approach}
\label{sec:result}

The optimization problem associated with the regularized least-squared based reconstruction problem involves minimizing the energy function $\E_0(\x)=\tfrac{(\y-\H\x )^2}{2 \sigma^2}+ V(\x)$. For the noise free case, using $\y = \H \x_0$, the energy to be optimized may be  rewritten as 
\begin{equation}
\E(\u) = \frac{1}{2\sigma^2} \u^\mathrm{T}\H^\mathrm{T}\H\u + V(\u+\x_0).
\end{equation}
where  $\u = \x - \x_0$. Note that, unlike the function $\E_0(\x)$, which is parametrized by known quantities (the data $\y$ and the measurement matrix $\H$) and can therefore be empirically optimized with respect to its argument, the closely related function $\E(\u)= \E_0(\u+\x_0)$ depends on the knowledge of the original signal $\x_0$.  The purpose of dealing with this function is {\it not} to provide an algorithm to estimate this signal given measured data, but to study the {\it statistical} behavior of this function and its minima over the distribution of problem instances, namely, input signals and the measurement matrices. For example, we can calculate the distribution of each component of the estimation error vector $\u$, given the distributions of $\x_0$ and $\H$. We will be working with $\E(u)$, although the susceptibility for a particular problem instance, to be defined below, could be defined completely in terms of $\E_0(\x)$.

In case this cost function reproduces the correct answer, the function $\E(\u)$ minimizes at $\u=0$. Looking at the structure of $\E(u)$ near zero tells us about potential shallow directions in parameter estimation error space, along which the cost function does rise significantly to reign in error. This failure could be quantified in terms of a susceptibility to error under linear perturbations to the cost function. The susceptibility provides a measure of robustness of the estimated parameters, and could indicate the trustworthiness of the reconstructed solution.

In addition, the cavity method derivation of the mean field equations involves considering  the effect of altering a single variable or of imposing an additional data constraint. These modifications would be treated as  `small' perturbations to the large-scale optimization problem. The susceptibility, especially the local part of it, would play a central role in computing the effect of these perturbations. 

\paragraph*{Definition of Susceptibility:}

As usual, let us consider a general regularization function $V(\x)$ for which there is a unique minimum to the cost function. Let the minimum of $\mathcal E(\u)$ be at $\u=\hat\u$. 
We introduce an augmented cost function 
\begin{equation}
\E(\u;\f) = \frac{1}{2\sigma^2} \u^\mathrm{T}\H^\mathrm{T}\H\u + V(\u+\x_0)-\f\cdot\u.
\end{equation}
with the variables $\f$, which are conjugate to $\u$. Optimizing $\E(\u;\f)$ will produce an $\f$ dependent answer $\u=\hat\u(\f)$. For small $\f$ we expect 
\begin{equation}
\hat\u(\f)=\hat\u+\CHI\f+\cdots
\label{eq:chidef}
\end{equation} 
defining the susceptibility matrix $\CHI$. Note that we want to take $M,N \rightarrow \infty$ first before we take the $\f \rightarrow 0$, to define susceptibility. We also expect that
\begin{equation}
\underset{\u}{\mathrm{min}}\!\ \mathcal E(\u;\f) =\mathcal E(\hat\u )-\hat\u^{\T}\f - \frac{1}{2}\f^{\T} \CHI \f +\cdots.
\label{eq:E-chi-expand}
\end{equation}

When $\H$ is a large random matrix, we can make asymptotic estimates of the mean and the variance of different components of the susceptibility matrix $\CHI$, following earlier work on singular values of random matrices~\cite{Anirvan99,Anirvan00}. This is carried out in Appendix \ref{app:susc}. Note that only the diagonal terms $\chi^{aa}$ have non-trivial means, whereas the off-diagonal terms average to zero. For diagonal terms, namely local susceptibilities, the average over all $a$'s,
 \begin{align}
\overline{\chi}\equiv\frac{1}{N}\sum_a\chi^{aa},
\label{eq:chibar_orig}
\end{align}
is expected to be self averaging in the large $M,N$ limit and be independent of the $\H$ for a matrix chosen from the distribution $\P(\H)$.

One should note,  $\chi^{ab}$, $a\neq b$,   is an  $\H,\x_0$-dependent number of the order $1/\sqrt{M}$ for a particular choice of $\H$ and of $\x_0$. It only vanishes after averaging over problem instances. Moreover,  even if these $O(1/\sqrt{M})$ terms are small compared to the $O(1)$ diagonal terms, the off-diagonal terms have an important effect on the self-consistency equations via the so-called Onsager reaction term~\cite{Onsager36}, as we will see in Appendix \ref{app:cav}. 

%

\paragraph*{Removing a Variable Node:}

\begin{figure}[t]
\begin{center}
\includegraphics[width=0.75\hsize]{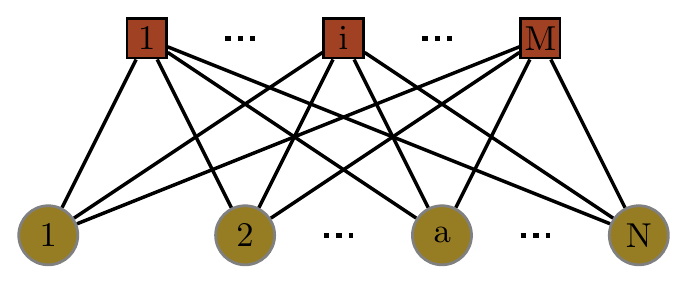}
\end{center}
\caption{Bipartite graph with variable nodes (circles) and constraint nodes (squares).}
\label{fig:bipartite}
\end{figure}

For the ensuing discussion, it is useful to visualize the problem in terms of a bipartite graph (see Fig.~\ref{fig:bipartite}), where the variables $x_a$ are represented by circular nodes and the `constraints' arising from each $y_i$ (namely, the terms $\tfrac{1}{2\sigma^2}(y_i-\sum_aH_{ia}x_a)^2=\tfrac{1}{2\sigma^2}(\sum_aH_{ia}u_a)^2$ in the cost function) are represented by  squares. Had we stuck to a finite temperature description, this graph would be the factor graph~\cite{Loeliger04}. If we insist on satisfying the condition $\y=\H\x$, this graph could be thought of as a Tanner graph~\cite{Tanner81}, with the circles being the  variable nodes and the squares being the `check' nodes. The system with $N$ variables (circles) and $M$ data constraints (squares) would be represented as the $(N,M)$ system. Our task is to relate properties of the $(N,M)$ system to $(N-1,M)$ system and obtain self-consistency conditions based on quantities that converge in the thermodynamic limit, $N,M\rightarrow \infty$.

We pick a particular node $a$ and partition the cost function 
\begin{equation}
\mathcal E(\u) = \frac{1}{2\sigma^2} \u^\mathrm{T}\H^\mathrm{T}\H\u + V(\u+\x_0)
\end{equation}
into a contribution purely from the node, a term representing the interaction of the node variable with the rest of the system, and, lastly, the cost function of the $(N-1,M)$ system:
\begin{align}
\mathcal E(\u) =& \frac{1}{2\sigma^2}u_a^2+ U(u_a+x_{0a}) + \frac{1}{\sigma^2} u_a \h_a \cdot \sum_{b\backslash a} \h_b u_b\nonumber\\
&+ \frac{1}{2\sigma^2} \big(\sum_{b\backslash a} \h_b u_b\big)^2 +\sum_{b\backslash a} U(u_b+x_{0b}).
\label{eq:E-expand}
\end{align}
Here, the $a$-th column of the $\H$ matrix is being represented by the vector $\h_a$, and, the subscript $\backslash a$ indicates that we leave out the node $a$. Moreover, we approximated $\h_a^2$ by its average value 
\begin{equation}
[\h_a^2]^{\mathrm{av}}_{\H}=\sum_i[H_{ia}^2]^{\mathrm{av}}_{\H}=\sum\limits_{i=1}^M\frac{1}{M}=1,
\end{equation}
since $\h_a^2$ is a sum of $M$ terms and is self-averaging. The typical fluctuation of  $\h_a^2$ from its average value $1$ asymptotically vanishes as $O(1/\sqrt{M})$. 

\begin{figure}[t]
\begin{center}
\includegraphics[width=0.75\hsize]{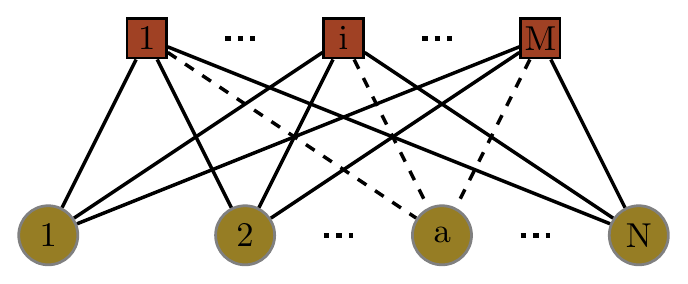}
\end{center}
\caption{The $(N-1,M)$ cavity system. Node $a$ has been removed from the system by removing the links to it.}
\label{fig:cavity-node}
\end{figure}

The system without node `$a$', i.e., the system with a `cavity' (see Fig.~\ref{fig:cavity-node}), will have its own optimum values  $u_b=\hat{u}_b$, for all $b\neq a$. The variable $u_a$ interacts with the rest of the system through the quantity  $\h_a\cdot\sum_{b\backslash a} \h_b u_b$. The program of cavity method is to characterize the distribution of this quantity in terms of some parameters relating to the $(N-1,M)$ system, and then use the fact that node `$a$' is statistically the same as every other node to relate these parameters to the distribution of $u_a$.

Since we are looking for the ground state, we minimize the expression in Eq. \eqref{eq:E-expand}. This optimization becomes equivalent to the minimization of the following system with the node variable, the Onsager reaction term~\cite{Onsager36} and a contribution from the system with the `cavity':
\begin{align}
\underset{\u}{\mathrm{min}}\!\ \mathcal E(\u)& =\underset{u_a}{\mathrm{min}}\!\ \{ \frac{1}{2\sigma^2}u_a^2
+ U(u_a+x_{0a})\nonumber\\
&-\frac{1}{2\sigma^2}\frac{\overline{\chi}}{\alpha\sigma^2+\overline{\chi}}u_a^2+\frac{1}{\sigma^2} u_a \h_a \cdot \sum_{b\neq a} \h_b \hat{u}_b + \mathcal E_{\backslash a}(\ubah ) \}.
\label{eq:E-couplings}
\end{align}
The detailed of this calculation is given in Appendix \ref{app:cav}.  The Onsager term $-\frac{1}{2\sigma^2}\frac{\overline{\chi}}{\alpha\sigma^2+\overline{\chi}}u_a^2$ appears as a reaction term toward the variable $u_a$ due to the adjustment of the other nodes after optimizing over them while holding $u_a$ fixed. The coefficient of $u_a^2$ turns out to be independent of $a$ for large systems, because of self-averaging. We combine the $u_a^2$ terms in Eq. \eqref{eq:E-couplings} to get
\begin{align}
\underset{\u}{\mathrm{min}}\!\ \mathcal E(\u)& =\underset{u_a}{\mathrm{min}}\!\ \{ \frac{1}{2\seff}u_a^2
+ U(u_a+x_{0a})\nonumber\\
&+\frac{1}{\sigma^2} u_a \h_a \cdot \sum_{b\neq a} \h_b \hat{u}_b + \mathcal E_{\backslash a}(\ubah ) \}
\label{eq:E-couplings-2}
\end{align}
with $\seff=\sigma^2+\overline{\chi}/\alpha$.

If we did not have $u_a$, the system would be optimized at $\ubah$. But now that $u_a$ is coupled to $\h_a \cdot \sum_{b\neq a} \h_b \hat{u}_b$, one needs to characterize the distribution of this quantity.  Let us define
\begin{equation}
\eta_a=-\h_a \cdot \sum_{b\neq a} \h_b \hat{u}_b=-\sum_{i=i}^M H_{ia}\hat v_i
\end{equation}
where $\hat v_i$ are the components of the residual vector $\hat \v=\sum_{b\neq a} \h_b \hat{u}_b$.  Note that $\h_a$ and $\v$ are independent variables. Over the ensemble of problem instances, $\eta_a$ has a distribution which is expected to be Gaussian, given that it is  a sum of many contributions. All we need to do is to calculate the mean and the variance of this variable, using the independence of $H_{ia}$'s and $v_j$'s.
\begin{align}
[\eta_a]^{\mathrm{av}}_{\x_0,\H}&=-\sum_i [H_{ia}]^{\mathrm{av}}_{\H}[\hat v_i]^{\mathrm{av}}_{\x_0,\H}=0\nonumber\\
[\eta_a^2]^{\mathrm{av}}_{\x_0,\H}&=\sum_{i,j} [H_{ia}H_{ja}]^{\mathrm{av}}_{\H}[\hat v_i\hat v_j]^{\mathrm{av}}_{\x_0,\H}\nonumber\\
&=\frac{1}{M}\sum_i [\hat v_i^2]^{\mathrm{av}}_{\x_0,\H}.
\label{eq:eta-moments}
\end{align}
Thus, all we need to know is the variance, $[\hat v_i^2]^{\mathrm{av}}_{\x_0,\H}$, of individual residuals for the $(N-1,M)$ system.

\paragraph*{Removing a Constraint Node:}

An individual component $\hat v_i$, being a sum of many variables, is expected to be Gaussian over the problem instance ensemble. The mean turns out to be zero but the variance requires a more careful analysis.   Because the variables  $H_{ib}$'s and $\hat u_c$'s are strongly correlated,
$[\hat v_i^2]^{\mathrm{av}}_{\x_0,\H}\neq \sum_{b\neq a} [H_{ib}^2]^{\mathrm{av}}_{\H} [\hat u_b^2]^{\mathrm{av}}_{\x_0,\H}$. To overcome this difficulty we need to replace $\hat u_b$'s by variables that are independent of the $i$-th row of the $\H$ matrix. A similar problem arises when using cavity method  in the context of Hopfield neural networks~\cite{Sompolinsky00}. 

In order to find these quantities, we go one step further  by removing the constraint `$i$'. The components of the error vector $u^{\prime}_b$ in the $(N-1,M-1)$ system  is indeed independent of the $i$-th row of the $\H$ matrix.    What remains to be done is to relate $\hat v_i$ to $u^{\prime}_b$'s.

To find $\hat v_i=  \sum_{b\neq a} H_{ib} \hat u_b$, we break up the minimization over $\uba$ into two steps:
\begin{equation}
\underset{\uba}{\mathrm{min}}\!\ \mathcal E_{\backslash a}(\uba) = \underset{  v_i}{\mathrm{min}}\bigg\{ \!\underset{ \begin{subarray}{c}
  \uba \\
 \mathrm{ s.t. }\sum\limits_{b\neq a} H_{ib}  u_b =v_i 
  \end{subarray} }{\mathrm{min}} \!\{ \mathcal E_{\backslash ai}(\uba) \}
+ \frac{1}{2\sigma^2} v_i^2 \bigg\}.
\label{eq:minEve}
\end{equation} 
\begin{figure}[t]
\begin{center}
\includegraphics[width=0.75\hsize]{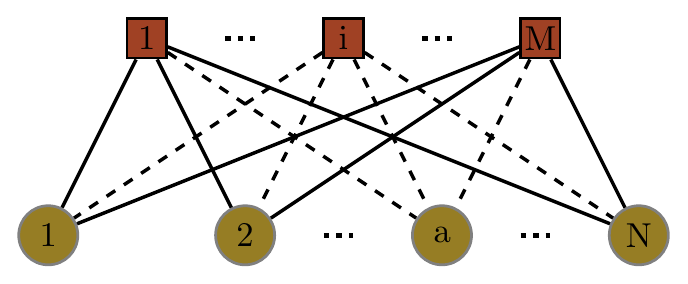}
\end{center}
\caption{The $(N-1,M-1)$ cavity system. Node $a$ and constraint $i$ have been removed from the system by removing the links to them.}
\label{fig:cavity-constraint}
\end{figure}
In Appendix~\ref{app:cav} we show that optimization of Eq. \eqref{eq:minEve} is equivalent to 
\begin{equation}
\underset{v_i}{\mathrm{min}}\!\ \{  \frac{\alpha}{2\overline{\chi}} (v_i-v^{\prime}_i )^2 +\frac{1}{2\sigma^2}v_i^2  \},\textrm{ with }   v^{\prime}_i=\sum\limits_{b\neq a}H_{ib} u^\prime_b.
\end{equation}
We expect the first term to be minimized at $v_i=v^{\prime}_i$, since minimization of the cost function  $\mathcal E_{\backslash ai}(\uba)$, without the $i$-th constraint, would be at $\uba=\uba^{\prime}$, making $v_i = \sum_{b\neq a} H_{ib}  u_b =\sum_{b\neq a}H_{ib} u^\prime_b=v^{\prime}_i$ at that point. The coefficient of $(v_i-v^{\prime}_i)^2$ depends only on $\overline{\chi}$ and not on $i$, because of self-averaging once again. The second term forces the residual to be small, imposing correlations between $H_{ib}$'s and $u_b$'s. We can capture the effect of such correlations by considering how $v_i$ is optimized when both terms are present.

Minimizing with respect to $v_i$ gives us
\begin{equation}
\hat v_i = \frac{1}{1+\frac{\overline{\chi}}{\alpha \sigma^2}}\sum\limits_{b\neq a} H_{ib} u^\prime_b.
\end{equation}
The denominator $(1+\frac{\overline{\chi}}{\alpha \sigma^2})$ `scales down' the unconstrained  answer $\sum_{b\neq a}H_{ib} u^\prime_b$. It is the same factor that relates $\sigma^2$ to $\seff$.

Given that this result is true for any $i$'s, Eq.~\eqref{eq:E-couplings} becomes
\begin{equation}
\underset{\u}{\mathrm{min}}\!\ \mathcal E(\u) =\underset{u_a}{\mathrm{min}}\!\ \{ \frac{1}{2\seff}u_a^2 -\frac{1}{\seff} \xi_a u_a+ U(u_a+x_{0a}) \}
\end{equation}
with 
\begin{align}
\xi_a \equiv -\sum_i H_{ia} \sum\limits_{b\neq a} H_{ib} u^{\prime}_b
\end{align}
being a random Gaussian variable with mean zero and variance $\sigma_\xi^2\equiv[\xi_a^2]^{\mathrm{av}}_{\x_0,\H}=\frac{q}{\alpha}$. The quantity $q \equiv \frac{1}{N-1} \sum_{b,c\neq a} [u^{\prime2}]^{\mathrm{av}}_{\x_0,\H}$ is the MSE  for the $(N-1,M-1)$ system. Insisting that $q$ is also the MSE of the $(N,M)$ system is one of the self-consistency conditions.

\paragraph*{Cavity Method Self-consistent Equations:}

In summary, the zero temperature problem boils down to optimizing a collection of `independent' variables
\begin{equation}
\hat u_a(f_a)=\underset{u_a}{\mathrm{argmin}}\!\ \{ \frac{1}{2\sigma_\mathrm{ eff}^2}(u_a^2- 2\xi_a u_a) + U(u_a+x_{0a})-f_au_a \}
\label{eq:min-singleE}
\end{equation}
using the same effective cost function as Eq.~\eqref{eq:Eeff-replica} in Sec.~\ref{sec:replica}, but with an additional linear perturbation. The variables $\xi_a$ are chosen independently from $\mathcal N(0,\sigma_\xi^2)$ and $x_0$'s are chosen independently from the probability distribution $p_0(x_0)$. With $x_{0a},\xi_a$ chosen randomly, we can obtain the distributions of $u_a(0)$ and $\chi^{aa}=\frac{d\hat u_a(f_a)}{df_a}\Big|_{f_a=0}$. 

The two parameters $\seff$ and $\sxi$  are decided by the following set of self-consistency conditions:
\begin{equation}
q = [  u_a(0)^2]^{\mathrm{av}}_{x_0,\xi}, \quad
\overline{\chi}=[ \chi^{aa}]^{\mathrm{av}}_{x_0,\xi}
\label{eq:self-consistency-cav}
\end{equation}
and
\begin{equation}
\seff = \sigma^2+\frac{\overline{\chi}}{\alpha}, \quad
\sxi=\frac{q}{\alpha}.
\label{eq:sigeff-cav}
\end{equation}

So far, we have analyzed the $\sigma_\zeta=0$ case. The presence of additive noise can be handled easily by our method, with $\hat v_i\equiv \sum_{b\neq a}H_{ib}\hat u_b+\zeta_i$ and $v^\prime_i\equiv \sum_{b\neq a}H_{ib} u^\prime_b+\zeta_i$. The effect of the additive noise could be absorbed in the $\xi_a$ variable, with the new self-consistency condition for the variance $\sxi$ would being: 
\begin{equation}
\sxi=\frac{q}{\alpha}+\sigma^2_{\zeta}.
\label{eq:sxi-zeta-cav}
\end{equation}

In this formulation, we do not need to invoke temperature. The average local susceptibility $\overline{\chi}$ plays the role of the quantity $\beta\Delta Q$  in the replica approach. In our subsequent work \cite{RMSTransition}, we will use $\overline{\chi}$ to distinguish phases around zero-temperature critical point, including the Donoho-Tanner transition~\cite{DonohoPT}.


\section{Mean Field Theory and Finite Size $\ell_1$-penalty Reconstruction}\label{sec:numeric}

The mean field self-consistency equations derived above are for the thermodynamic limit, namely for $M,N\to\infty$.
Many of the the quantities we define, like    $\overline{\chi}$ and $\seff (=\sigma^2+\overline{\chi}/\alpha)$, strictly make sense only after we take the problem size to infinity. How do these concepts arise in finite sized problems? To investigate this question, we look at  the average local susceptibility $\overline{\chi}$ for the important case of Basis Pursuit, which corresponds to $V(\x)=||\x||_1$ and to $\sigma\to 0$.
\begin{figure}[t]
\begin{center}
\includegraphics[width=.9\hsize]{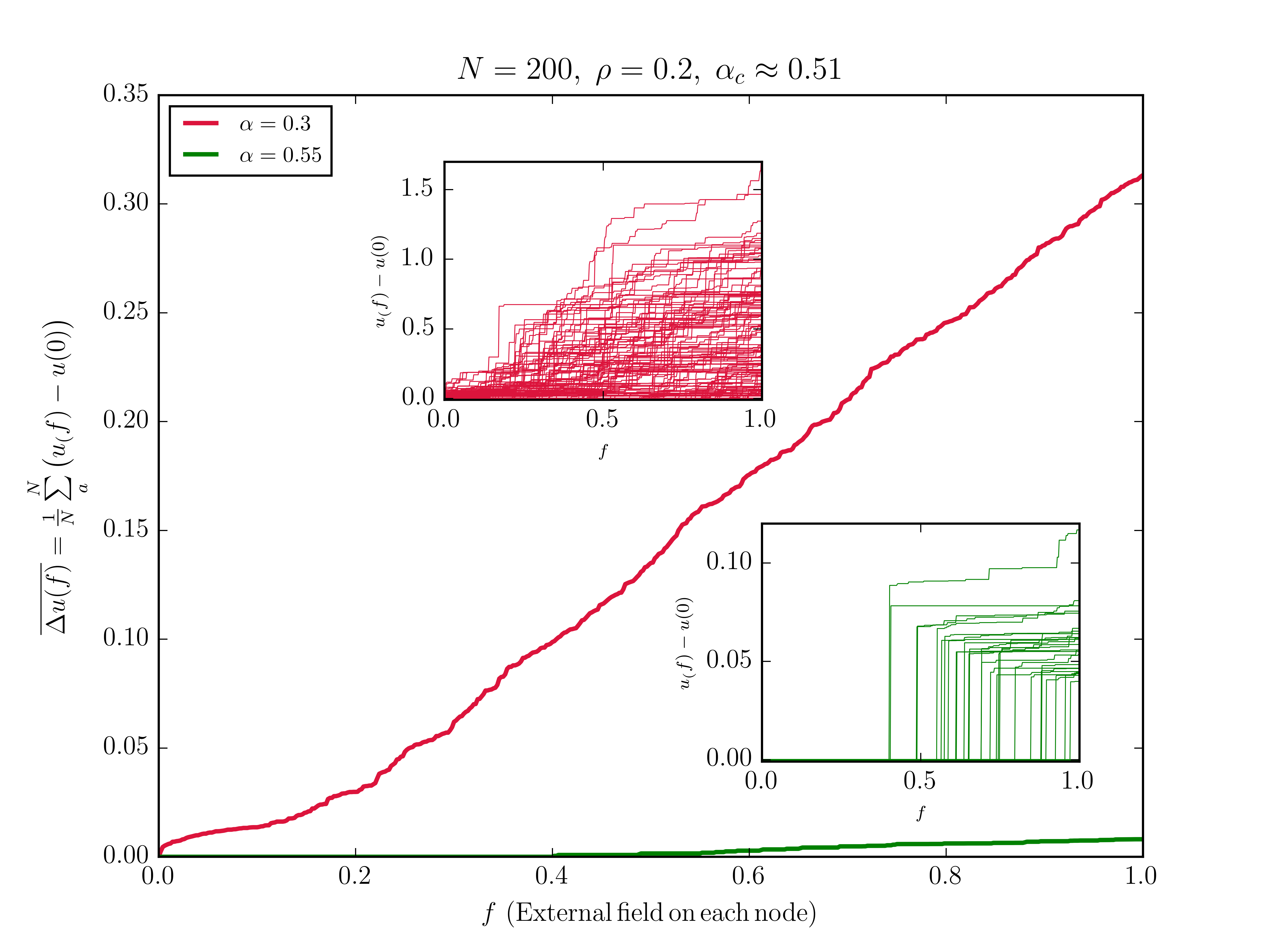} 
  \end{center}
\caption{Plot of the average error of the solution as a function of  external field `$f$' on each node in two different regimes: green line  being for perfect recovery regime and the red line corresponding error-prone one. The insets with corresponding colors are the responses of individual nodes, showing the staircase like behavior mentioned in the text. }
\label{fig:susc}
\end{figure}

In particular, we carry out the numerical experiment for minimization of $  ||\u+\x_0||_1-fu_a$, for each $a=1,\ldots,N$, subject to $\H\u=\bm0$, by linear programming using the CVXOPT package~\cite{Andersen10cvxopt}. We obtain the $M\times N$ matrix $\H$ by filling it with independent entries from a Gaussian distribution with mean zero and variance $1/M$. In this example, the size of the vector $\x$ is $N = 200$, and contains $K=40$ randomly placed elements driven from a standard Gaussian distribution. 

Since we solve a linear programming problem with the cost function perturbed by the linear term $-fu_a$, solutions are chosen from the extreme points of a convex polytope. As $f$ is changed, the solution, would jump from one extreme point to another at particular thresholds. As a result, functions $u_a(f)$ look like a set of staircases (see insets in Fig.~\ref{fig:susc}). 

To see the average local susceptibility emerge in the thermodynamic limit from these step functions, we need to compute the average response. The average parameter estimation error, $\overline{\Delta u(f)}=\tfrac{1}{N}\sum_a(u_a(f)-u_a(0))$, as a function of  external field `$f$', shown for different regimes in Fig.~\ref{fig:susc}, are strikingly different. In the high $\alpha$  solution, the average error has no response to the external field up to a large threshold. However, in the low $\alpha$ case, very small external fields can perturb the estimated solution to a new one. This is an indication of the robustness of the solution in the good recovery region and lack of it in the error-prone regime, and hints at a phase transition in between. In this particular case, that transition takes place at $\alpha=\alpha_c\approx0.51$.

To connect results shown in Fig.~\ref{fig:susc} to the average local susceptibility, we need to make linear fit to the average response $\overline{\Delta u(f)}$ as a function of $f$ near $f=0$. Note that differentiating $u_a(f)$ with respect to $f$ for the finite size system and then summing over $a$ would give us a very different answer.   In the good regime, the average local susceptibility, obtained by the fit, is approximately zero. As one decreases the number of constraints on the system past a threshold, this susceptibility becomes non-zero, as can be seen from the slope of the average response near zero, for the low $\alpha$ solution in Fig.~\ref{fig:susc}.


\section{Afterword}

In this study, we directly treat the regularized least-squared optimization problem and show how to adapt the cavity method for doing mean field theory in the context. The mean field theory leads to a self-consistency condition on average mean squared error (MSE), since error in estimating one variable affects error in others. Careful derivation of the self-consistency condition, without using replica trick, involve accounting for subtle correlations in the system. To take care of these correlations, we needed a two-step cavity approach: one step removing a variable and then, another, removing a data constraint.

Although, we have emphasized the zero-temperature treatment, the cavity method can be used for finite temperature results as well. For completeness, we have provided the corresponding derivation in Appendix~\ref{app:A}. The key connection with the zero-temperature treatment is via the fluctuation-dissipation theorem~\cite{Kubo66}, which relates thermal fluctuation with susceptibility.

The cavity approach looks at the behavior of the system for a particular choice of quenched variables, $\H$ and $\x_0$ in this case. In contrast, the replica approach centers on immediately averaging those quenched variables away. In the context of compressed sensing, one can imagine many problems, where the matrix $\H$ is non-random. Currently there  is no obvious way to extend the replica mean field treatment for such sensing matrices. The cavity method could be a more versatile tool in this regard. Extensions of this tool to other classes of compressed sensing problems would be a goal of future studies. 

\appendix

\section{Susceptibility and Conjugate Fields}\label{app:susc}
We want to minimize the augmented cost function 
\begin{equation}
\E(\u;\f) = \frac{1}{2\sigma^2} \u^\mathrm{T}\H^\mathrm{T}\H\u + V(\u+\x_0)-\f\cdot\u.
\end{equation}
producing an $\f$-dependent minimum $\u=\hat\u(\f)$. The susceptibility matrix $\CHI$ is defined by the  small $\f$ expansion: $\hat\u(\f)=\hat\u(0)+\CHI\f+\cdots$. If $\mathcal E$ is differentiable, the optimum $\tilde\u(\f)$ is the solution of
\begin{equation}
\f = \nabla_{\u} \mathcal E(\u)
\label{eq:fT}
\end{equation}
Where $\f=\mathbf 0$, $\u$ is at its optimal value $\hat\u$. Formally, if perturbation $\f$ is small and $\E$ is differentiable to higher orders, we can expect $\delta \u$ to be small and, therefore, Taylor expand $\mathcal E(\u + \delta \u)$ around $\u=\hat\u$
\begin{equation}
\mathcal E(\hat\u + \delta \u) = \mathcal E(\hat\u )
+  \frac{1}{2} \sum_{ab}\delta u_a\delta u_b \frac{\partial^2 \mathcal E}{\partial u_a \partial u_b}\bigg|_{\u=\hat\u} +\cdots.
\label{eq:dualE-expan}
\end{equation}
From \eqref{eq:fT} and \eqref{eq:dualE-expan}, we can identify the inverse susceptibility
$(\CHI^{-1})_{ab} =\tfrac{\partial^2 \mathcal E}{\partial u_a \partial u_b}\big|_{\u=\hat\u} $ 
and can show that
\begin{equation}
\underset{\u}{\mathrm{min}}\!\ \mathcal E(\u;\f) =\mathcal E(\hat\u )-\hat\u^{\T}\f - \frac{1}{2}\f^{\T} \CHI \f +\cdots.
\label{eq:E-chi-expand-app}
\end{equation}

It is simplest to study the properties of $\CHI$, when the potential $U(x)$ has continuous second derivatives. 
If we Taylor expand around the solution $\u=\hat\u$, we will have
 \begin{align}
\E(\hat\u+\delta\u;\f) = &\E(\hat\u;0)
+\frac{1}{2} \delta\u^\mathrm{T}[\frac{\H^\mathrm{T}\H}{\sigma^2}+\W(\x)]\delta\u \nonumber\\&-\f\cdot(\hat\u+\delta\u)+\cdots.
\end{align}
where $W_{ab}(\x)=U''(\hat{u}_{a}+x_{0a})\delta_{ab}$. Optimizing over $\delta \u$, we see that the susceptibility matrix would be given by
 \begin{equation}
\CHI(\x,\H)=[\frac{\H^\mathrm{T}\H}{\sigma^2}+\W(\x)]^{-1}.
\label{eq:smooth-case-susc}
\end{equation}

These statements are true, for fixed $N$, with $\f$ and $\delta\u$ going to zero. However, we are interested in the opposite limit, $N\to \infty$ first and then taking $\f,\delta\u$ small. We also need to deal with $U(x)$ that is singular. The way we will treat this difficulty is as follows. We will keep $U(x)$ to be smooth, take $N\to\infty$ limit on $\CHI$ defined by Eq.~\ref{eq:smooth-case-susc}, with the assumption that $W_{aa}(\x)$'s come from a well-defined distribution. In that case, when $\H$ is a large random matrix, we can make asymptotic estimates of the mean and the variance of different components of the susceptibility matrix $\CHI$, using diagrammatic expansions and resumming. 

\begin{figure}[t]
\begin{center}
\includegraphics[width=1.\hsize]{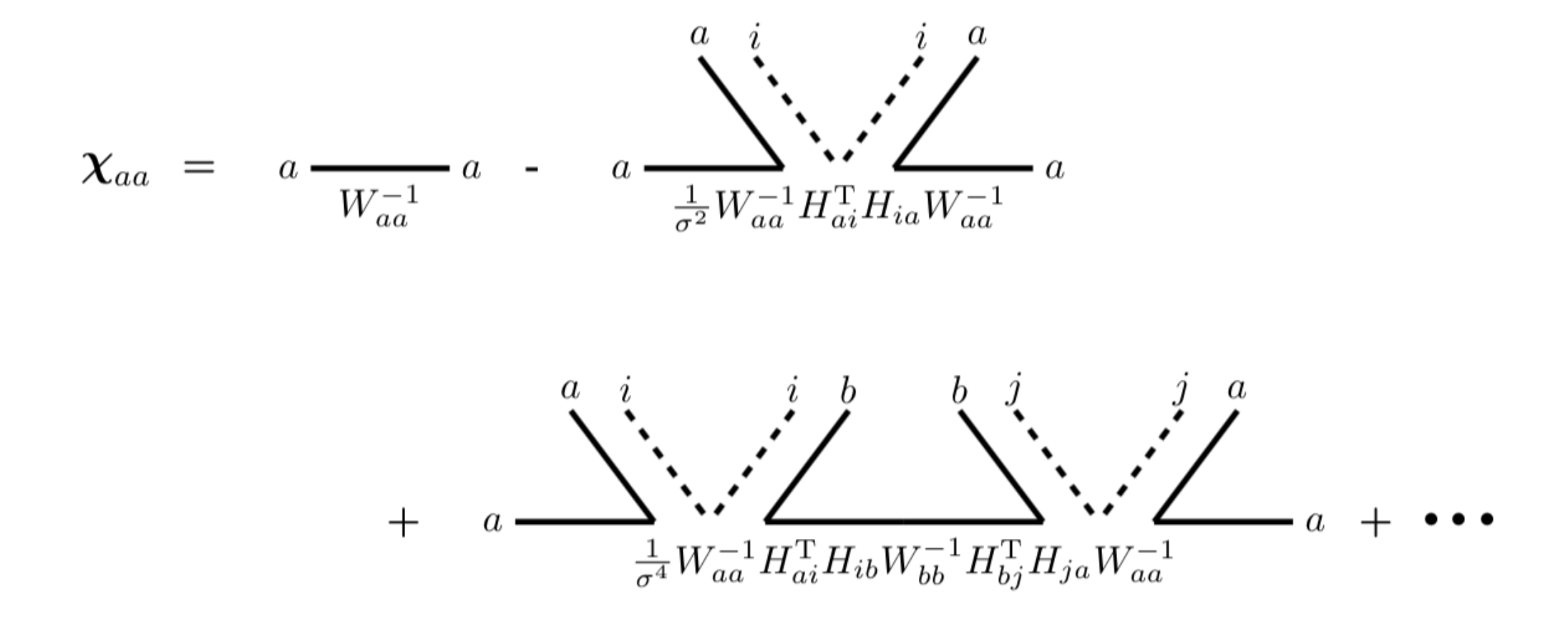}
\end{center}
\caption{The diagrammatic expansion of susceptibility.}
\label{fig:chi}
\end{figure}

\begin{figure}[t]
\begin{center}
\includegraphics[width=1.\hsize]{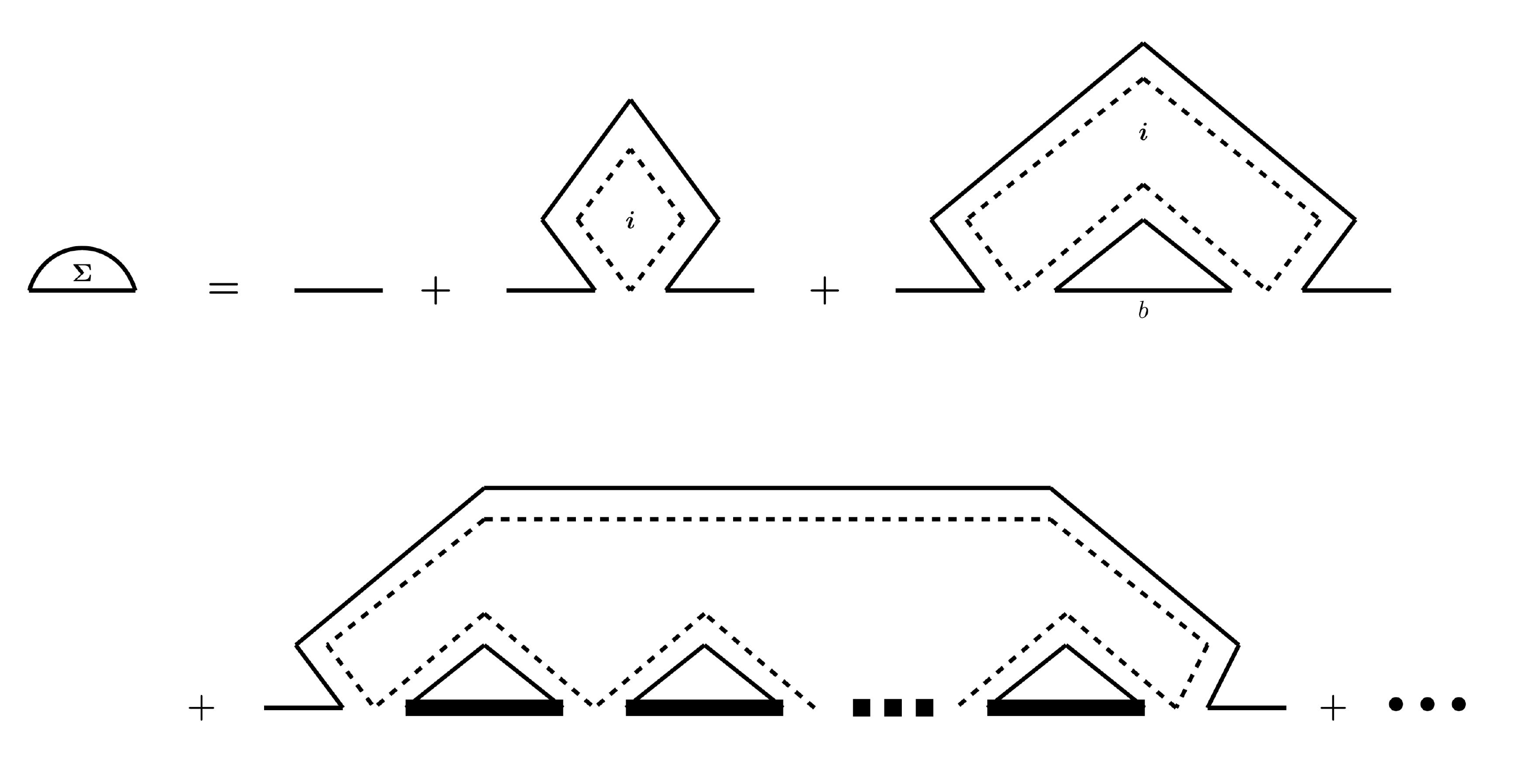}
\end{center}
\caption{Planar diagrams contributing to the self-energy.}
\label{fig:selfenergy}
\end{figure}

\begin{figure}[t]
\begin{center}
\includegraphics[width=1.0\hsize]{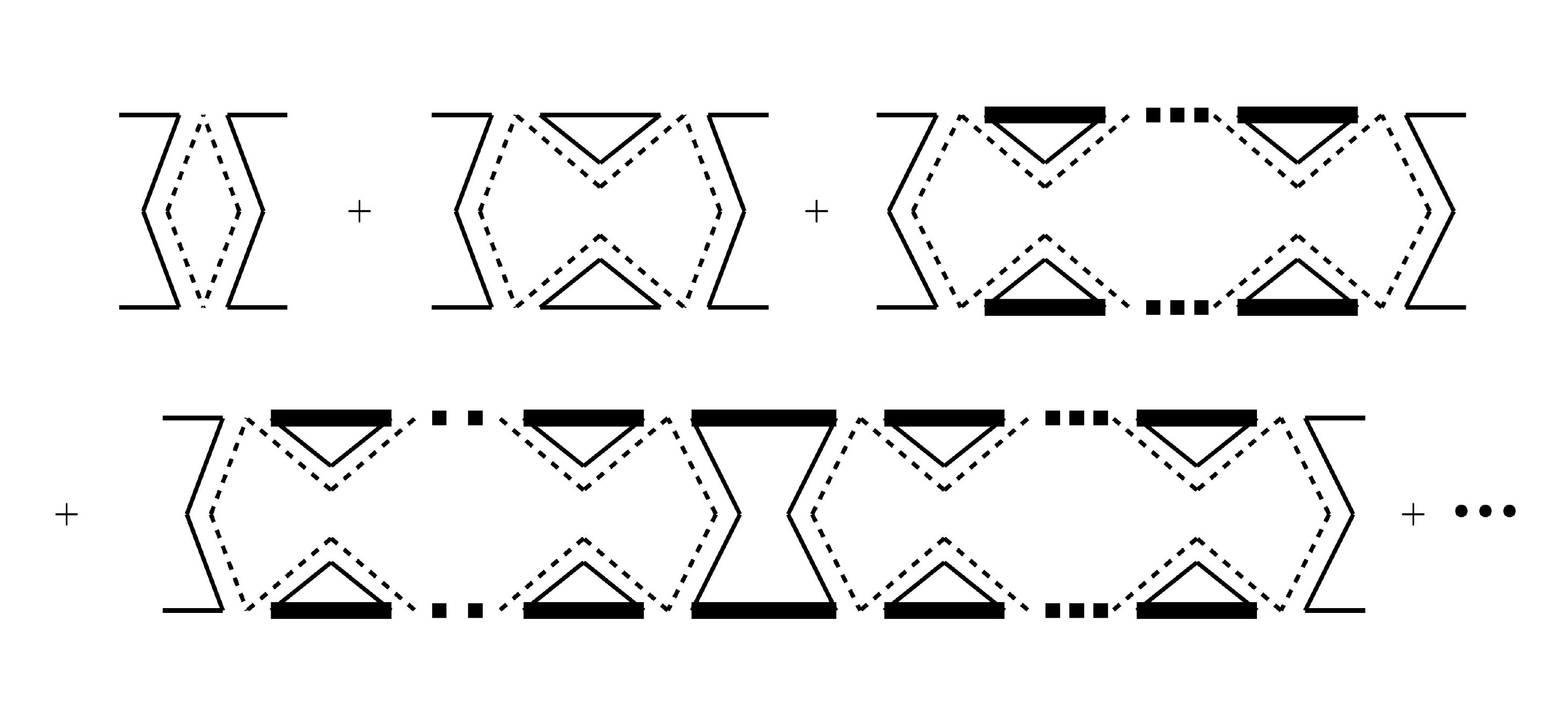}
\end{center}
\caption{The leading planar diagrams in covariance computation are of the order $O(\tfrac{1}{M},\tfrac{1}{N})$, as can be seen from counting a factor of $M$ or $N$ for appropriate index loop, and counting a factor of $\tfrac{1}{M}$ for each double-line contraction coming from averaging over the matrix elements. }
\label{fig:chichi}
\end{figure}

To find the asymptotic behavior of $\CHI$, we formally expand the RHS of Eq.~\eqref{eq:smooth-case-susc} in powers of $\tfrac{\H^\mathrm{T}\H}{\sigma^2}$ (see Fig.~\ref{fig:chi}) and compute moments by averaging over $H_{ia}$ diagrammatically. Namely, we expand
 \begin{align}
\CHI=&\W^{-1}-\frac{1}{\sigma^2}\W^{-1} \H^\mathrm{T}\H\W^{-1}\nonumber\\
&+\frac{1}{\sigma^4}\W^{-1}\H^\mathrm{T}\H\W^{-1}\H^\mathrm{T}\H\W^{-1}-\cdots
\label{eq:smooth-case-susc-expand}
\end{align}
and then compute moments of the form $[\chi^{a_1b_1}\chi^{a_2b_2}\cdots \chi^{a_kb_k}H_{i_1c_1}\cdots H_{i_lc_l}]^{\mathrm{av}}_{\H}$ using Wick's theorem~\cite{peskin1995introduction}, since $\H$ distribution is Gaussian with mean and covariance specified by 
Eq.~\eqref{eq:gauss-av} and Eq.~\eqref{eq:gauss-covar}, respectively.  
One can write $[\CHI(\x,\H)]^{\mathrm{av}}_{\H}$ as $[\W(\x)-\Sigma(\x)\mathbf{I}_N]^{-1}$, where $\Sigma(\x)$ is a self-energy term.  Using the fact that, in the large $M,N$ limit, only the planar diagrams survive, the contributions to the self-energy are shown in Fig.~\ref{fig:selfenergy} and can be re-summed as
\begin{equation}
\Sigma(\x)=-\frac{1}{\sigma^2}\frac{1}{1+\frac{1}{M\sigma^2}\sum_a\chi^{aa}(\x)}
\end{equation}
Hence, the mean susceptibility (holding $x_a$'s fixed but averaging over $\H$) is given by
 \begin{equation}
 \CHI^{\mathrm{av}}(\x)\equiv
[\CHI(\x,\H)]^{\mathrm{av}}_{\H}=\bigg[\W(\x)+\frac{M}{M\sigma^2+\mathrm{Tr}[\CHI^{\mathrm{av}}(\x)]}\mathbf{I}_N\bigg]^{-1}.
\label{eq:smooth-case-susc-av}
\end{equation}
Covariance of $\CHI$ entries, $[\chi^{ab}(\x,\H)\chi^{cd}(\x,\H)]^{\mathrm{av}}_{\H}$
 could be computed using the diagrams in Fig.~\ref{fig:chichi} and  they are suppressed in the large $M, N$ limit, since their contributions are $O(\tfrac{1}{M},\tfrac{1}{N})$.

Therefore, we get the local susceptibilities, i.e. diagonal terms, to be:
 \begin{align}
&[\chi^{aa}(\x)]^{\mathrm{av}}_{\H}=\bigg [W_{aa}(x_a)+\frac{1}{\sigma^2+\frac{\overline{\chi}(\x)}{\alpha}}\bigg]^{-1}
\label{eq:uncorr-susc-av}
\\
&\overline{\chi}(\x)\equiv\frac{1}{N}\sum_a[\chi^{aa}(\x)]^{\mathrm{av}}_{\H}.
\label{eq:chibar}
\end{align}

Here, $[\chi^{ab}]^{\mathrm{av}}_{\H}=0$ for $a\neq b$, but each $\chi^{ab}$ is of order $O(1/\sqrt{M})$. In particular, we will need the correlation of $\chi^{ab}$ with the corresponding matrix elements of $\H^\mathrm{T}\H$. 
Using the identity $[\W+\tfrac{\H^\mathrm{T}\H}{\sigma^2}]\CHI=\mathbf{I}_N$, we can prove a useful corollary of the  result in Eq.~\eqref{eq:smooth-case-susc-av}.
 \begin{align}
&\frac{1}{\sigma^2}[\H^\mathrm{T}\H\CHI(\x,\H)]^{\mathrm{av}}_{\H}=\mathbf{I}_N-\W(\x)\CHI^{\mathrm{av}}(\x)\nonumber\\
&=\mathbf{I}_N-\W(\x)[\W(\x)+\frac{M}{M\sigma^2+\mathrm{Tr}[\CHI^{\mathrm{av}}(\x)]}\mathbf{I}_N]^{-1}\nonumber\\
&=\frac{M}{M\sigma^2+\mathrm{Tr}[\CHI^{\mathrm{av}}(\x)]}[\W+\frac{M}{M\sigma^2+\mathrm{Tr}[\CHI^{\mathrm{av}}(\x)]}\mathbf{I}_N]^{-1}\nonumber\\
&=\frac{M\CHI^{\mathrm{av}}(\x)}{M\sigma^2+\mathrm{Tr}[\CHI^{\mathrm{av}}(\x)]}\nonumber\\
&=\frac{\alpha\CHI^{\mathrm{av}}(\x)}{\alpha\sigma^2+\overline{\chi}(\x)}
\label{eq:smooth-case-susc-hth-av}
\end{align}
In particular, Eq.~\eqref{eq:smooth-case-susc-hth-av}
implies
\begin{equation}
[\mathrm{Tr}(\H^\mathrm{T}\H\CHI(\x,\H))]^{\mathrm{av}}_{\H}=\frac{M\sigma^2\overline{\chi}(\x)}{\alpha\sigma^2+\overline{\chi}(\x)}
\label{eq:smooth-case-susc-hth--uncorr-av}
\end{equation}
which we will use this identity in Appendix \ref{app:cav}.

 Notice that many observations made here are independent of the assumption that $U(x)$ has a continuous second derivative. For example, in the case of compressed sensing with $U(x)=\lambda|x|$, we could define a second-differentiable function $U_\epsilon(x)$ such that $\lim_{\epsilon\rightarrow 0}U_\epsilon(x)=U(x)$, for example, $U_\epsilon(x)=\sqrt{x^2+\epsilon^2}$ or $U_\epsilon(x)=\tfrac {1}{\epsilon}\log(2\cosh( \epsilon x))$.  If $x_a=x_{0a}+u_a$ goes to zero as $\epsilon$ vanishes, then the corresponding $W_{aa}=U_\epsilon''(x_a)$  diverges. However, the corresponding local susceptibility, $\chi^{aa}$, just becomes zero in this limit. Therefore,  as $\epsilon\rightarrow0$, the idea of using effective single variable optimization problems and determining the self-consistent distribution of $x_a$ and $\chi^{aa}$ remains meaningful. We just need to separate out the set of variables $x_a$ for which $W_{aa}$ diverges and treat this set carefully.  As a consequence of $\CHI$ remaining well-defined in the $\epsilon\rightarrow 0$ limit.


\section{Zero-temperature Cavity Method}\label{app:cav}

%

\subsection{Removing a Variable Node}

We expand the following equation in terms of node `$a$'
\begin{equation}
\mathcal E(\u) = \frac{1}{2\sigma^2} \u^\mathrm{T}\H^\mathrm{T}\H\u + V(\u+\x_0)
\end{equation}
which results in 
\begin{align}
\mathcal E(\u) =& \frac{1}{2\sigma^2}u_a^2+ U(u_a+x_{0a}) + \frac{1}{\sigma^2} u_a \h_a \cdot \sum_{b\backslash a} \h_b u_b\nonumber\\
&+ \frac{1}{2\sigma^2} \big(\sum_{b\backslash a} \h_b u_b\big)^2 +\sum_{b\backslash a} U(u_b+x_{0b})
\label{eq:E-expanded}
\end{align}
The system with a cavity (at node `$a$') has a new optimum values $u_b=\hat{u}_b$, for all $b\neq a$. In the complete system, the variable from node `$a$', namely $u_a$, interacts with the rest of the variable via the term $\h_a\cdot\sum_{b\backslash a} \h_b u_b$. We rewrite Eq.~\eqref{eq:E-expanded} representing the the interaction between the node and the rest as a perturbation by a field:
\begin{equation}
\mathcal E(\u) =  \frac{1}{2\sigma^2}u_a^2+ U(u_a+x_{0a}) +\mathcal E_{\backslash a}(\u_{\backslash a})  -\u_{\backslash a}^{\mathrm T}\f_{\backslash a}. 
\label{eq:Ef}
\end{equation}
The cost function of the $(N-1,M)$ system is $\mathcal E_{\backslash a}(\u_{\backslash a})$. We identify $(\f_{\backslash a})_b=-\frac{1}{\sigma^2}\h_b\cdot \h_a u_a$ to be the local force exerting on each node $u_b$, due to presence of node $u_a$. Since we are looking for the ground state, we minimize the expression in Eq.~\eqref{eq:Ef} 
\begin{align}
\underset{\u}{\mathrm{min}}\!\ \mathcal E(\u) &=\underset{u_a,\u_{\backslash a}}{\mathrm{min}}\{ \frac{1}{2\sigma^2}u_a^2+ U(u_a+x_{0a}) \nonumber\\
&+ \mathcal E_{\backslash a}(\u_{\backslash a}) -\u_{\backslash a} ^{\mathrm T}\f_{\backslash a}\}.
\label{eq:OEba}
\end{align}

Given that $\h_b\cdot \h_a$ is of the order $1/\sqrt{M}$, $\f_{\backslash a}$ is small, and we can invoke the definition of susceptibility $\CHI_{\backslash a}$ for the $(N-1,M)$ system and use expansion of the minimized cost function~\eqref{eq:E-chi-expand-app}.
\begin{align}
\underset{\u}{\mathrm{min}}\!\ \mathcal E(\u) &=\underset{u_a}{\mathrm{min}}\!\ \{ \frac{1}{2\sigma^2}u_a^2+ U(u_a+x_{0a})\nonumber\\
&+ \mathcal E_{\backslash a}(\ubah )-\ubah^{\T}\f_{\backslash a} - \frac{1}{2}\f_{\backslash a}^{\T} \chiba \f_{\backslash a} \}
\end{align}
and plugging in $(\f_{\backslash a})_b=-\frac{1}{\sigma^2}\h_b\cdot \h_a u_a$, we get
\begin{align}
\underset{\u}{\mathrm{min}}\!\ \mathcal E(\u)& =\underset{u_a}{\mathrm{min}}\!\ \{ \frac{1}{2\sigma_\mathrm{ eff}^2}u_a^2
+ U(u_a+x_{0a})\nonumber\\
&+\frac{1}{\sigma^2} u_a \h_a \cdot \sum_{b\neq a} \h_b \hat{u}_b + \mathcal E_{\backslash a}(\ubah ) \}
\label{eq:E-coupling}
\end{align}
where
\begin{equation}
\frac{1}{\sigma_\mathrm{ eff}^2}=\frac{1}{\sigma^2}\left(1- \frac{1}{\sigma^2}\sum_{b,c\neq a} (\h_a\cdot \h_b)(\h_a\cdot \h_c) \chi_{\backslash a}^{b c} \right) 
\label{eq:sigmaeff2}
\end{equation}
The quantity $\sum_{b,c\neq a} ( \h_b)_i( \h_c)_j \chi_{\backslash a}^{b c}$, is independent of $\h_a$. As a result, $\sum_{b,c\neq a} (\h_a\cdot \h_b)(\h_a\cdot \h_c) \chi_{\backslash a}^{b c}$ can be replaced by  $\sum_{b,c\neq a} ( \h_b)\cdot( \h_c) \chi_{\backslash a}^{b c}/M$, thanks to the self-averaging of $( \h_a)_i( \h_a)_j$. Using Eq.~\eqref{eq:smooth-case-susc-hth--uncorr-av} for the $(N-1,M)$ system,
\begin{equation} 
\sum_{b,c\neq a}  \h_b\cdot \h_c \chi_{\backslash a}^{b c}
=\frac{M\sigma^2\overline{\chi}_{\backslash a}}{\alpha\sigma^2+\overline{\chi}_{\backslash a}}
\approx \frac{M\sigma^2\overline{\chi}}{\alpha\sigma^2+\overline{\chi}}
\label{eq:barchi-identity}
\end{equation}
with the last step having to do with $\overline{\chi}$ becoming independent of $N,M$ asymptotically. Using this last relation Eq.~\eqref{eq:barchi-identity}, in Eq.~\eqref{eq:sigmaeff2} we get
\begin{equation}
\sigma_\mathrm{ eff}^2=\sigma^2+\frac{\overline{\chi}}{\alpha}
\label{eq:sigmaeff2-final}
\end{equation}

Looking at Eq.~\eqref{eq:E-coupling}, we still have to determine the size of the coupling of the node variable $u_a$ to the rest of the system via $\eta_a = -\h_a\cdot\hat \v$ with $\hat \v=\sum_{b\neq a} \h_b \hat{u}_b$. In Section \ref{sec:result}, we showed that the first and the second moments of the $\eta_a$ are:
\begin{align}
[\eta_a]^{\mathrm{av}}_{\x_0,\H}=&0\nonumber\\
[\eta_a^2]^{\mathrm{av}}_{\x_0,\H}=&\frac{1}{M}\sum_i[\hat v_i^2]^{\mathrm{av}}_{\x_0,\H}.
\end{align}
The order $k$ cumulants go as $M^{1-k/2}$ and, for $k>2$, they tend to zero as $M$ goes to infinity.   Therefore, we will stop with the variance and treat $\eta_a$ as a zero-mean Gaussian variable.  We still need the variance, for which we need a condition to determine $[\hat v_i^2]^{\mathrm{av}}_{\x_0,\H}$. This requires a second step of the cavity method.
\subsection{Removing a Constraint Node}

The subtlety in determining $[\hat v_i^2]^{\mathrm{av}}_{\x_0,\H}$ involves accounting for correlation between matrix elements $H_{ib}$ and the optimal values $\hat{u}_b$ of the $(N-1, M)$ system. To do this, we need to set up an $(N-1,M-1)$ system with the constraint `$i$' removed (see Fig.~\ref{fig:cavity-constraint}). To find $\hat v_i=  \sum\limits_{b\neq a} H_{ib} \hat u_b$, we write the minimization over $\uba$ as follows:
\begin{equation}
\underset{\uba}{\mathrm{min}}\!\ \mathcal E_{\backslash a}(\uba) = \underset{  v_i}{\mathrm{min}}\bigg\{ \!\underset{ \begin{subarray}{c}
  \uba \\
 \mathrm{ s.t. }\sum\limits_{b\neq a} H_{ib}  u_b =v_i 
  \end{subarray} }{\mathrm{min}} \!\{ \mathcal E_{\backslash ai}(\uba) \}
+ \frac{1}{2\sigma^2} v_i^2 \bigg\}
\label{eq:minEv}
\end{equation} 
with the first minimization being a constrained one for the $(N-1,M-1)$ system, subject to $\sum_{b\neq a} H_{ib}  u_b =v_i$, and the second one being over $v_i$. The cost function for the system without nodes $a,i$ is represented by $\mathcal E(\uba)_{\backslash i}$. The term $\tfrac{1}{2\sigma^2} v_i^2$ represents the constraint coming from the $i$-th observation. Had we done an unconstrained optimization of $\mathcal E_{\backslash ai}(\uba)$, the optimum $\ubah$ would be independent of $H_{ib}$. Trying to keep $v_i$ small perturbs this solution by a small amount and induces correlation with $H_{ib}$. Our strategy would be to compute the effect of  perturbation in terms of the system susceptibility.

In order to do constrained minimization, we use the Lagrange multiplier method
\begin{align}
\underset{\uba}{\mathrm{min}}\!\ \mathcal E_{\backslash a}(\uba) = &\underset{\gamma_i}{\mathrm{max}}\underset{\uba, v_i}{\mathrm{min}}\!\ \{ \mathcal E_{\backslash ai}(\uba) \nonumber\\
&+ \frac{1}{2\sigma^2} v_i^2 - \gamma_i (v_i - \sum\limits_{b\neq a} H_{ib}  u_b ) \}.
\label{eq:minEv-Lagrange}
\end{align}

 Minimizing Eq.~\eqref{eq:minEv-Lagrange} with respect to $v_i$ we get $v_i=\sigma^2\gamma_i$, and making that substitution for $v_i$ into the cost function we get
\begin{align}
\underset{\uba}{\mathrm{min}}\!\ \mathcal E_{\backslash a}(\uba) =&\underset{\gamma_i}{\mathrm{max}} \,\underset{\uba}{\mathrm{min}}\!\ \{ \mathcal E_{\backslash a}(\uba) - 
\frac{1}{2}\sigma^2 \gamma_i^2 -\uba^{\T}\g  \}\\
=& \underset{\gamma_i}{\mathrm{max}}\!\ \{  - \frac{1}{2}\sigma^2 \gamma_i^2 +\mathcal E^{*}_{\backslash i}(\g) \}
\label{eq:minEgamma}
\end{align}
with $g_b = -\gamma_i H_{ib}$ and with
 $\mathcal E^{*}_{\backslash i}(\g)$  is defined as
\begin{equation}
\mathcal E^{*}_{\backslash i}(\g)= \underset{\uba}{\mathrm{min}}\!\ \{ \mathcal E_{\backslash ai}(\uba)  -\uba^{\T}\g \}
\end{equation}
where the  presence of $\g$  alters the optimal $\uba$ from the unconstrained optimum $\ubap$. Since each component of $\g$ is small ($O(1/\sqrt{M})$), we can expand around $\ubap$ using $\chibai$, the susceptibility of the $(N-1, M-1)$ system, as in Eq.~\eqref{eq:E-chi-expand-app}.  Therefore, $\mathcal E^{*}_{\backslash i}(\g)$ can be written as
\begin{equation}
\mathcal E^{*}_{\backslash i}(\g)=\mathcal E_{\backslash i}(\ubap ) -\uba^{\prime\T}\g - \frac{1}{2}\g^{\T} \chibai \g+\cdots
\end{equation}
Now, Eq.~\eqref{eq:minEgamma} becomes
\begin{equation}
\underset{\uba}{\mathrm{min}}\!\ \mathcal E_{\backslash a}(\uba) =\underset{\gamma_i}{\mathrm{min}}\!\ \{ - \frac{1}{2}\sigma^2 \gamma_i^2 + 
\mathcal E_{\backslash ai}(\ubap ) -\uba^{\prime\T}\g - \frac{1}{2}\g^{\T} \chibai \g \}
\end{equation}

The quadratic term $\g^{\T} \chibai \g=\gamma_i^2\sum_{ij}H_{ib}H_{ic}\chi^{bc}_{\backslash ai}$ can be simplified because of self-averaging. We have
\begin{equation}
\sum_{ij}[H_{ib}H_{ic}]^{\mathrm{av}}_{\H}\chi^{bc}_{\backslash ai}=\frac{1}{M}\sum_b\chi^{bb}_{\backslash ai}\approx\frac{\overline{\chi}}{\alpha},
\end{equation}
once more using the fact that the average local susceptibility $\overline{\chi}$ is nearly the same for the $(N,M)$ system and the $(N-1,M-1)$ system.

Putting everything together
\begin{equation}
\underset{\uba}{\mathrm{min}}\!\ \mathcal E_{\backslash a}(\uba) =\underset{\gamma_i}{\mathrm{max}}\!\ \{  - \frac{\sigma^2}{2} (1+\frac{\overline{\chi}}{\alpha \sigma^2} ) \gamma_i^2 + 
\gamma_i \sum\limits_{b\neq a} H_{ib} u^\prime_b \},
\end{equation}
maximizing with respect to $\gamma_i$ and then using $v_i=\sigma^2\gamma_i$ gives us
\begin{equation}
v_i = \frac{1}{1+\frac{\overline{\chi}}{\alpha \sigma^2}}\sum\limits_{b\neq a} H_{ib} u^\prime_b.
\end{equation}

Since this result is true for any $i$'s, Eq. \eqref{eq:E-coupling} becomes
\begin{equation}
\underset{\u}{\mathrm{min}}\!\ \mathcal E(\u) =\underset{u_a}{\mathrm{min}}\!\ \{ \frac{1}{2\sigma_\mathrm{ eff}^2}u_a^2- 
\frac{\xi}{\sigma^2(1+\frac{\overline{\chi}}{\alpha \sigma^2})}u_a + U(u_a+x_{0a}) \}
\end{equation}
with 
\begin{equation}
\xi \equiv -\sum_i H_{ia} \sum\limits_{b\neq a} H_{ib} u^{\prime}_b
\end{equation}
being a random Gaussian variable with mean zero and variance
\begin{align}
\sigma_\xi^2\equiv[\xi^2]^{\mathrm{av}}_{\x_0,\H}&= \sum_{i,j}[H_{ia} H_{ja}]^{\mathrm{av}}_{\H} \sum\limits_{b,c\neq a}[H_{ib} H_{jc}]^{\mathrm{av}}_{\H}\; [u^\prime_b u^\prime_c]^{\mathrm{av}}_{\x_0,\H}\nonumber\\
&=\sum_{i,j}\frac{\delta_{ij}}{M} \sum\limits_{b,c\neq a}\frac{\delta_{ij}\delta_{bc}}{M}\; [u^\prime_b u^\prime_c]^{\mathrm{av}}_{\x_0,\H}\nonumber\\
&=\frac{1}{M} \sum\limits_{b\neq a}\; [u^{\prime 2}_b]^{\mathrm{av}}_{\x_0,\H}=\frac{q}{\alpha}
\label{eq:sigxi}
\end{align}
thanks to $H_{ja}$ and $H_{ib}$ being independent for $a\neq b$, as well as $u^\prime_b$'s being independent of those matrix elements.
Here the quantity $q \equiv \frac{1}{N-1} \sum\limits_{b,c\neq a} [u^{\prime2}]^{\mathrm{av}}_{\x_0,\H}$ is the MSE for the $(N-1,M-1)$ system. Taking into consider that $q$ is a self-averaging quantity and is asymptotically the same for the $(N,M)$ system, we get the independent single variable optimization  
\begin{equation}
\underset{u_a}{\mathrm{min}}\!\ \{ \frac{1}{2\sigma_\mathrm{ eff}^2}(u_a^2- 2\xi_au_a) + U(u_a+x_{0a}) \}
\label{eq:minEeff}
\end{equation}
with $\xi_a \in \mathcal N(0,\sigma_\xi^2)$ and $\seff=\sigma^2+\overline{\chi}/\alpha$.

\section{Finite Temperature Cavity Method}\label{app:A}

In this section, we solve the finite temperature problem formulated in Sec.~\ref{sec:reconstruction} via the cavity method.
With the cost function written in terms of $\u$ as
\begin{equation}
\E(\u) =\frac{1}{2\sigma^2}(\H \u)^{2} + V(\u+\x_0)
\label{eq:E}
\end{equation}
we define the Boltzmann distribution $P(\u|\H,\x_0)$:
\begin{equation}
P(\u|\H,\x_0)=\frac{1}{Z(\beta|\H,\x_0)}e^{- \beta \E}
\label{eq:BoltzmannDist}
\end{equation}
with the normalization factor/partition function given by
\begin{equation}
Z(\beta|\H,\x_0)  = \int d\u \,e^{- \beta \E} 
\end{equation}

We now apply the first step of the two-step cavity method. First, we rewrite $\E$ as 
an interaction between variable $u_a$ and the rest of the variables
\begin{equation}
\E(\u) = \frac{1}{2\sigma^2} \h_{a}^2 u_a^2 + \frac{1}{\sigma^2} u_a \h_a\cdot \sum_{b \neq a} \h_b u_b+ U(u_a+x_{0a}) + \E_{\backslash a} (\uba)
\end{equation}
By defining
\begin{equation}
\eta_a \equiv -\frac{ \h_a\cdot \sum_{b \neq a} \h_b u_b}{\h_a^2}
\end{equation}
and using $\h_{a}^2=1+O(\tfrac{1}{\sqrt{M}})$ we have
\begin{equation}
\E = \frac{1}{2\sigma^2} ( u_a^2 -2 u_a\eta_a)+ U(u_a+x_{0a}) + \E_{\backslash a}(\uba)
\label{eq:Eba}
\end{equation}
with $\uba$, $\E_{\backslash a}$ etc defined as in Sec.~\ref{sec:result}. 
Equation~\eqref{eq:Eba} indicates that the variable $u_a$ interacts with all the others 
only through $\eta_a$. Therefore, we rewrite the marginal distribution $P(u_a)$ as an integral over the joint distribution of $\eta_a$ and $u_a$, $P(u_a,\eta_a)$.
\begin{equation}
 P(u_a) = \frac{1}{Z} \int d\uba\, \mathrm{e}^{-\beta \E} =  \int d\eta_a \,P(u_a,\eta_a)
\label{eq:Pmarginal}
\end{equation}
 where
\begin{equation}
P(u_a,\eta_a)=\frac{1}{Z}\int d \uba\, \delta(\eta_a+ \h_a\cdot \sum_{b \neq a} \h_b u_b)\,\mathrm{e}^{-\beta \E}\,
\label{eq:Pjoint}
\end{equation}
for all $a=1,\ldots, N$.
Now we introduce a cavity `field' distribution of $\eta_a$ at the removed node $a$ as
\begin{equation}
 P_{\backslash a}(\eta_a)=\frac{1}{Z_{\backslash a}}\int d \u \;\delta(\eta_a+ \h_a\cdot \sum_{b \neq a} \h_b u_b)\mathrm{e}^{-\beta \E_{\backslash a}}.
\label{eq:Pcavity}
\end{equation}
By comparing \eqref{eq:Pjoint} and \eqref{eq:Pcavity}, we get
\begin{equation}
 P(u_a) = \tfrac{\int d \eta_a \exp\big[-\beta\big\{\frac{( u_a^2 -2 u_a\eta_a)}{2\sigma^2} + U(u_a+x_{0a})\big\} \big] P_{\backslash a}(\eta_a)}{\int d u_a d \eta_a 
\mathrm\exp\big[-\beta\big\{\frac{( u_a^2 -2 u_a\eta_a)}{2\sigma^2} + U(u_a+x_{0a})\big\} \big]P_{\backslash a}(\eta_a)}
\label{eq:PmarginalCavity}
\end{equation}

The assumption of continuity of the global ground state, even in the presence of the cavity after removing node $a$, is equivalent to the replica symmetric hypothesis. This is a valid assumption when the penalty function $V$ 
is convex. Therefore, in the limit of $N \rightarrow \infty$, 
even if the nodes of $(N-1,M)$ system are weakly correlated,  $\eta_a$ is still a sum of many variables and  $P(\eta_a)_{\backslash a}$ can well be approximated by a 
Gaussian distribution.

\begin{equation}
 P_{\backslash a}(\eta_a)\propto e^{-\frac{(\eta_a - \langle\eta_a\rangle_{\backslash a})^2}{2 \langle\delta \eta_a^2\rangle_{\backslash a}}}
\end{equation}
Then \eqref{eq:PmarginalCavity} becomes
\begin{equation}
 P(u_a)= \tfrac{ \exp\{-\frac{\beta}{2 \sigma^2} \left( 1 - \frac{\beta}{\sigma^2}\langle\delta \eta_{a}^2\rangle_{\backslash a} \right) u_a^2 + 
\frac{\beta u_a}{\sigma^2}\langle\eta_a\rangle_{\backslash a}  -\beta U(u_a+x_{0a})\}}{ \int du_a \exp\{-\frac{\beta}{2 \sigma^2} \left( 1 - \frac{\beta }{\sigma^2}\langle\delta \eta_a^2\rangle_{\backslash a} \right) u_a^2 + 
\frac{\beta u_a}{\sigma^2}\langle\eta_a\rangle_{\backslash a}  -\beta U(u_a+x_{0a})\}}
\label{eq:Pmarginal1step}
\end{equation}
Therefore, only the thermal averages $\langle\eta_a\rangle_{\backslash a}$ and the thermal fluctuation strength $\langle\delta \eta_a^2\rangle_{\backslash a}=\langle(\eta_a-\langle\eta_a\rangle_{\backslash a})^2\rangle_{\backslash a}$ of the field $\eta_a$ for the distribution $P_{\backslash a}(\eta_a)$ are left to be computed. 
In that process the  effects of (weak) correlation between the $u_a$'s have to be accounted for. Define, as in Sec.~\ref{sec:result},
\begin{equation}
v_{i} = \sum_{b\neq a} H_{ib} u_b
\end{equation}
and utilize our definition,
\begin{equation}
\eta_a = - \sum_{i} H_{ia} v_i
\end{equation}
then we arrive at
\begin{equation}
\langle \eta_a \rangle_{\backslash a} =   - \sum_{i} H_{ia} \langle v_i\rangle
\label{eq:eta}
\end{equation}
and 
\begin{align}
\langle \delta \eta_a^2 \rangle_{\backslash a} &= \sum_{ij} H_{ia} H_{ja}\langle\delta v_i\delta v_j\rangle\nonumber\\
&\approx\sum_{ij} \frac{1}{M}\delta_{ij} \langle\delta v_i\delta v_j\rangle =\frac{1}{M}\sum_i\langle\delta v_i^2\rangle
\label{eq:eta2}
\end{align}
Having done that we need to compute $\langle v_i\rangle$ and $\langle\delta v_i^2\rangle$. To do so, this time in addition to site $a$ we exclude site $i$. Hence from \eqref{eq:Eba} we get
\begin{equation}
\mathcal{E}_{\backslash a}(\uba) = \frac{1}{2\sigma^2} v_i^2 + \mathcal{E}_{\backslash ai}(\uba)
\end{equation}
After carrying out the same computation as in \eqref{eq:Pmarginal}, \eqref{eq:Pjoint}, and \eqref{eq:PmarginalCavity} for the marginal distribution $Q_{\backslash a}(v_i)$, we arrive at
\begin{equation}
Q_{\backslash a}(v_i)= \frac{  \exp\big\{-\frac{\beta}{2 \sigma^2} v_i^2 -\frac{\left( v_i - \langle v_i\rangle_{\backslash ai}\right)^2}{2 \langle\delta v_i^2\rangle_{\backslash ai}}\big\}}
{\int d v_i\exp\big\{-\frac{\beta}{2 \sigma^2} v_i^2 -\frac{\left( v_i - \langle v_i\rangle_{\backslash ai}\right)^2}{2 \langle\delta v_i^2\rangle_{\backslash ai}}\big\}}
\end{equation}
Therefore 
\begin{equation}
Q_{\backslash a}(v_i)= \frac{  \exp\bigg\{-\frac{\beta}{2 \sigma^2} \big(1+\frac{\sigma^2}{\beta \langle\delta v_i^2\rangle_{\backslash ai}}\big)
\bigg(v_i - \frac{ \langle v_i\rangle_{\backslash ai}}{ 1+ \frac{\beta \langle\delta v_i^2\rangle_{\backslash ai}}{\sigma^2}}\bigg)^2\bigg\}}
{\int d v_i \exp\bigg\{-\frac{\beta}{2 \sigma^2} \big(1+\frac{\sigma^2}{\beta \langle\delta v_i^2\rangle_{\backslash ai}}\big)
\bigg(v_i - \frac{ \langle v_i\rangle_{\backslash ai}}{ 1+ \frac{\beta \langle\delta v_i^2\rangle_{\backslash ai}}{\sigma^2}}\bigg)^2\bigg\}}
\end{equation}
and then $\langle\delta v_i^2 \rangle$ is 
\begin{equation}
\langle\delta v_i^2 \rangle = \frac{1}{\frac{\beta}{\sigma^2} (1+\frac{\sigma^2}{\beta \langle\delta v_i^2\rangle_{\backslash ai}})}
= \frac{\langle\delta v_i^2\rangle_{\backslash ai}}{1+ \frac{\beta \langle\delta v_i^2\rangle_{\backslash ai}}{\sigma^2}}
\label{eq:vi2}
\end{equation}
and $\langle v_i\rangle$ is at 
\begin{equation}
\langle v_i\rangle =  \frac{ \langle v_i\rangle_{\backslash ai}}{1+ \frac{\beta \langle\delta v_i^2\rangle_{\backslash ai}}{\sigma^2}}.
\label{eq:vi}
\end{equation}
Notice how both these moments for the $(N-1,M)$ system is scaled down by the same factor, when compared to the moments for the $(N-1,M-1)$ system. Using arguments similar to the fluctuation-dissipation~\cite{Kubo66} theorem, we could show that the change in $\langle v_i\rangle$ due a change in $\langle v_i\rangle_{\backslash ai}$, susceptibility of sorts, is closely related to $\langle\delta v_i^2\rangle_{\backslash ai}^{-2}$ times  $\langle\delta v_i^2\rangle$, with the first term of the product playing the role of temperature.

Carrying on, we get
\begin{align}
\langle \delta v_i^2 \rangle_{\backslash ai} &= \sum_{b,c\neq a} H_{ib} H_{ic} \langle \delta u_b \delta u_c \rangle_{\backslash ai}\nonumber\\
&=\sum_{b,c\neq a} \frac{1}{M}\delta_{bc} \langle \delta u_b \delta u_c \rangle_{\backslash ai}+ O(\frac{N}{M^{3/2}},\frac{N^{1/2}}{M})\nonumber\\
&\approx =\frac{1}{M}\sum_{b\neq a} \langle \delta u_b^2 \rangle_{\backslash ai}
\end{align}
since the $(N-1,M-1)$ system, indicated by the subscript `${\backslash ai}$', is independent of  $H_{ib}$ and $H_{ic}$,  $H_{ib} H_{ic}=\tfrac{1}{M}\delta_{bc}+O(\tfrac{1}{M})$ fluctuations,  and $\langle \delta u_b \delta u_c \rangle_{\backslash ai}\sim O(\tfrac{1}{\sqrt{M}},\tfrac{1}{\sqrt{N}})$ when $b\neq c$, indicating that nodes are only weakly correlated.

To make connection with the notation in Sec.~\ref{sec:replica}, let us introduce $\Delta Q$
\begin{equation}
\Delta Q \equiv \frac{1}{N}\sum_{a} \langle \delta u_a^2\rangle
\approx \frac{1}{N-1}\sum_{b\neq a} \langle \delta u_b^2\rangle_{\backslash ai},
\label{eq:DQ-app}
\end{equation}
the second approximate equality becoming exact in the thermodynamic limit.
Then, we have
\begin{equation}
\langle \delta v_i^2 \rangle_{\backslash ai} = \Delta Q/\alpha.
\end{equation}

Therefore from \eqref{eq:eta2}, \eqref{eq:vi2}, and \eqref{eq:DQ-app} 
\begin{equation}
\frac{\beta}{\sigma^2}\langle \delta \eta_{a}^2 \rangle_{\backslash a} = \frac{1}{ (1+\frac{\sigma^2}{\beta \Delta Q/\alpha})}
\label{eq:meta2}
\end{equation}
and from \eqref{eq:eta}, \eqref{eq:vi}, and \eqref{eq:DQ-app}
\begin{equation}
\langle \eta_{a} \rangle_{\backslash a} =  \frac{\sum_i H_{ia} \sum_{b\neq a} H_{ib} \langle u_b\rangle_{\backslash ai}}{\big(1 + \frac{\beta \Delta Q}{\alpha \sigma^2}\big)}.
\label{eq:meta}
\end{equation}
Moreover, we define
\begin{equation}
\xi_a \equiv \sum_i H_{ia} \sum_{b\neq a} H_{ib} \langle u_b\rangle_{\backslash ai}
\label{eq:xi}
\end{equation}
which has variance $\sigma_\xi^2= q/\alpha$ with $q$
\begin{equation}
q = \frac{1}{N}\sum_{a} \langle u_a\rangle^2\approx\frac{1}{N-1}\sum_{b} \langle u_b\rangle^2_{\backslash ai}
\label{eq:mse-q}
\end{equation}
being the mean squared error. Therefore, by plugging \eqref{eq:meta} and \eqref{eq:xi} into Eq.~\eqref{eq:Pmarginal1step}, the marginal distribution for single variable $u_a$ becomes
\begin{equation}
P(u_a)= \frac{ \exp\{-\frac{\beta }{2 \sigma_{\mathrm{eff}}^2}( u_a^2 -2u_a\xi_a) -\beta U(x_{0a}+u_a)\}}
{ \int du_a \exp\{-\frac{\beta }{2 \sigma_{\mathrm{eff}}^2}( u_a^2 -2u_a\xi_a) -\beta U(x_{0a}+u_a)\}}
\label{eq:one-node-dist}
\end{equation}
with $\sigma_{\mathrm{eff}}^2 = \sigma^2(1+\frac{\beta\Delta Q}{\alpha \sigma^2})$, and the effective cost function for the individual node is
\begin{equation}
\E(u_a) = \frac{1}{2 \sigma_{\mathrm{eff}}^2}( u_a^2 -2u_a\xi_a) +U(x_{0a}+u_a)
\label{eq:Eeffc}
\end{equation}
Therefore, with $\E$ replaced by a set of effectively decoupled nodes, and the sum over index $a$ replaced by a quenched average over $\xi_a,x_{0a}$. As a result, the self-consistency conditions for the MSE
\begin{equation}
q = \frac{1}{N}\sum _{a=1}^N\langle u_a \rangle^2\\
\end{equation}
and for
\begin{equation}
\Delta Q = \frac{1}{N}\sum_{a=1}^N \langle \delta u_a^2 \rangle
\end{equation}
reduce to
\begin{equation}
q = \big[\langle u \rangle_{\mathrm{eff}}^2\big]^{\mathrm{av}}_{\xi,x_0}
\end{equation}
and
\begin{equation}
\Delta Q = \big[\langle \delta u^2 \rangle_{\mathrm{eff}}\big]^{\mathrm{av}}_{\xi,x_0}
\label{eq:DQ-app-2}
\end{equation}
where the thermal average $\langle \ldots\rangle_\mathrm{eff}$ is performed with respect to the effective individual node distribution~\eqref{eq:prob-eff} and $\big[\ldots\big]^{\mathrm{av}}_{\xi,x_0}$ is the quenched average over variables $\xi,x_0$, with $\xi$ drawn from  ${\cal N}(0,q/\alpha)$ and signal $x_0$ drawn independently from a distribution $P(x_0)$. These self-consistency equations are exactly the same those from the replica symmetric ansatz in Sec.~\ref{sec:replica}.

\begin{acknowledgments}
M.R. and A.M.S. acknowledge the hospitality of the Simons Center for Data Analysis. A.M.S. thanks Pankaj Mehta for comments on an earlier version of the manuscript. This work was supported by the National Science Foundation INSPIRE (track 1) award 1344069 to P.P.M. and A.M.S.

\end{acknowledgments}

\bibliographystyle{apsrev}

\bibliography{CS-references}

\end{document}